\documentclass[reprint,twocolumn,aps,pra,longbibliography,superscriptaddress,nofootinbib,10pt]{revtex4-2}

\usepackage[english]{babel} 	
\usepackage[usenames, dvipsnames]{xcolor} 
\usepackage{soul}

\usepackage{graphicx} 
\usepackage{bm} 
\usepackage{amsmath} 
\usepackage{amsthm} 
\usepackage{amssymb} 
\usepackage{times} 
\usepackage[pdftex,colorlinks=true,urlcolor= blue,linkcolor=Blue,citecolor=RedViolet]{hyperref}
\usepackage{multirow}

\usepackage{physics}
\usepackage{appendix} 

\usepackage{enumitem}

\newtheorem{theorem}{Theorem}
\newtheorem{corollary}{Corollary}
\newtheorem{proposition}{Proposition}

\usepackage{dsfont}

\providecommand{\bra}[1]{\langle #1|}
\providecommand{\ket}[1]{|#1 \rangle}

\providecommand{\ketbra}[2]{|#1\rangle \langle #2|}

\newcommand{\PRA}{Phys. Rev. A }

\newcommand{\PRL}{Phys. Rev. Lett. }

\newcommand{\mc}[1]{\mathcal{#1}}

\begin{document}

\title{A unified fidelity-based framework for quantum speed limits in open quantum systems}
\author{Tristán M. Osán}
\email{tristan.osan@unc.edu.ar}
\affiliation{Instituto de F\'{i}sica Enrique Gaviola (IFEG-UNC-CONICET), Universidad Nacional de Córdoba, 
Av. Medina Allende s/n, Ciudad Universitaria. (CP:X5000HUA) Córdoba. Argentina}
\affiliation{Universidad Nacional de Córdoba. Facultad de Matemática, Astronomía, Física y Computación. Grupo de teoría de la materia condensada. Av. Medina Allende s/n, Ciudad Universitaria. (CP:X5000HUA) Córdoba. Argentina}
\author{Yanet Alvarez}
\affiliation{Instituto de Física La Plata (IFLP, CONICET-UNLP). Diag. 113 entre 63 y 64. La Plata (CP:1900). Buenos Aires. Argentina.}
\affiliation{Universidad Nacional de La Plata. Facultad de Ciencias Exactas. Calle 115 y 47. La Plata (CP:1900). Buenos Aires. Argentina.}
\author{Mariela Portesi}
\affiliation{Instituto de Física La Plata (IFLP, CONICET-UNLP). Diag. 113 entre 63 y 64. La Plata (CP:1900). Buenos Aires. Argentina.}
\affiliation{Universidad Nacional de La Plata. Facultad de Ciencias Exactas. Calle 115 y 47. La Plata (CP:1900). Buenos Aires. Argentina.}
\author{Pedro W. Lamberti}
\affiliation{Universidad Nacional de Córdoba. Facultad de Matemática, Astronomía, Física y Computación. Grupo de teoría de la materia condensada. Av. Medina Allende s/n, Ciudad Universitaria. (CP:X5000HUA) Córdoba. Argentina}
\affiliation{Consejo Nacional de Investigaciones Científicas y Técnicas (CONICET). Av. Medina Allende s/n, Ciudad Universitaria. (CP:X5000HUA) Córdoba. Argentina.}

\begin{abstract}
In this work, we investigate the role of functionals of generalized fidelity measures in deriving quantum speed limits (QSLs) within a geometric approach.
We establish a general theoretical framework and show that, once a specific generalized fidelity is selected, the resulting Margolus–Levitin and Mandelstam–Tamm QSLs for both unitary and nonunitary (Lindblad-type) dynamics depend solely on the chosen fidelity measure.
We prove that any monotone, differentiable reparametrization of the chosen fidelity yields exactly the same Margolus–Levitin and Mandelstam–Tamm-type QSL, rendering the QSLs invariant under such transformations and thereby unifying fidelity- and metric-induced geometric formulations.
This result highlights the limitations of improving QSLs through functional transformations of fidelity and indicates that genuine improvements must arise from alternative fidelity definitions.
We further show that several QSLs reported in the literature are encompassed by our framework and discuss possible extensions based on other generalized fidelity measures beyond those explicitly analyzed.
In addition, we consider the damped Jaynes–Cummings model as a concrete physical setting to explore the behavior of the QSLs discussed in this work and analyze their regime-dependent features.
\end{abstract}

\maketitle


\section{Introduction}

A quantum computer, or any system that processes quantum information, may be viewed more fundamentally as a quantum system that evolves from a given initial state to a physically distinguishable final state under a prescribed set of operations. A natural requirement is that the rate of change of the quantum state should be as fast as possible, both to minimize the total processing time and to limit the exposure of the quantum system to uncontrollable interactions with its environment, which may lead to undesirable phenomena such as decoherence. Notably, quantum physics imposes fundamental limits on the rate at which a quantum state can evolve. These limits are known as quantum speed limits (QSLs). The problem of evaluating the QSL in systems evolving according to certain established dynamics has been a subject of research since the early years of quantum theory. For example, in the case of isolated quantum systems, quantum speed limits such as those of Mandelstam and Tamm (MT) and Margolus and Levitin (ML) set bounds on the minimum time required for a system to evolve from a given initial state to an orthogonal quantum state~\cite{Mandel1945, Margolus1998, Fleming1973, Bhatta1983, Anandan1990, Vaidman1991, Uffink1993, Brody2003}.
However, in realistic quantum processes one often aims not to reach an orthogonal state, but rather to drive the system toward an arbitrary, possibly non-orthogonal, target state. Motivated by this idea, several studies have extended the original QSL framework to account for unitary evolution between non-orthogonal states (see, for example, Refs.~\cite{Pfeifer1993,Giovannetti2003,Luo2004,Levitin2009,Morley-Short2014,Hornedal2023}).\par
Over the past decade, various approaches have extended QSLs beyond idealized closed dynamics to mixed states and open systems using, for example, similarity measures such as Bures fidelity, relative purity, and Fisher-information-based distances ~\cite{Deffner2013, delCampo2013, Campaioli2018, Taddei2013}. Each approach has illuminated a different aspect of the QSL landscape \cite{Campaioli2019, Funo2019, OConnor2021, Thakuria2024, Hornedal2025, Srivastav2025, Mai2023, Mai2025a, Mai2025b, delCampo2025, Mirkin2016}. However, these efforts typically fix a single distinguishability measure at the outset, leaving open the question of whether a unified principle underlies all fidelity-based QSLs.\par
The notion of a “geometric approach’’ appears frequently in the QSL literature, yet its meaning can vary substantially depending on the context and the mathematical tools being employed. While it is often associated with fidelity-based distance measures, it can also refer to more rigorous constructions rooted in differential geometry and information geometry~\cite{Jones2010,Zwierz2012,Deffner2013,Pires2016,Lan2022}. \par
An earlier and foundational example of geometric reasoning is due to Anandan and Aharonov~\cite{Anandan1990}, who derived a QSL for closed-system unitary evolution by considering the Fubini–Study distance between pure states. In this case, the speed limit arises naturally from the geometry of the projective Hilbert space and reflects the minimal time required for a quantum state to evolve into an orthogonal one~\cite{Anandan1990}.\par
In the work of Deffner and Lutz~\cite{Deffner2013}, the QSLs are framed in terms of the Bures angle, i.e., a functional of the Bures fidelity $F_B(\rho_0,\rho_t)$ between the initial and final quantum states $\rho_0$ and $\rho_t$, respectively \cite{bur69,Uhlm76,joz94}. Although their results are presented within a ``geometric'' perspective, no intrinsic metric properties of the Bures angle are used in the derivation. The essential role is played by the Bures fidelity $F_B(\rho_0,\rho_t)$ itself, together with standard functional inequalities. In this sense, the approach is better understood as a \emph{functional fidelity-based} method, and not a genuinely geometric derivation in the information‐geometric or Riemannian sense.\par
In contrast, Taddei et al.~\cite{Taddei2013} developed a geometric QSL by directly using the quantum Fisher information, which defines a Riemannian metric on the space of quantum states. Their approach compares the length of the actual evolution path to the geodesic distance in this metric, making essential use of differential-geometric concepts. This method reveals that the speed limit arises not just from functional properties of fidelity, but from the underlying geometry of state space itself. While this formulation has a deep conceptual significance, its practical implementation becomes challenging when the initial state is mixed, as the QSL requires a minimization of the quantum Fisher information over all possible purifications in an enlarged system–environment Hilbert space.\par
This line of reasoning was extended by Pires et al.~\cite{Pires2016}, who showed that many contractive Riemannian metrics—not just the Bures one—can generate valid QSLs. Their work demonstrates that one can construct an entire family of QSLs by choosing the appropriate metric structure, and that the tightest QSL may depend on the details of the physical dynamics.\par
From this landscape, it becomes clear that the phrase ``geometric approach'' can refer to rather different things: from a fidelity-based reformulation of dynamical bounds~\cite{Deffner2013}, to rigorous geometric derivations based on path length, metric contraction, and geodesics~\cite{Taddei2013,Pires2016}.\par

One central objective of this work is to reexamine the assumptions underlying the geometric formulation as employed by Deffner and Lutz in their seminal work~\cite{Deffner2013}, as well as related approaches proposed in the literature (see, for example, Refs.~\cite{Jones2010, Zwierz2012, Wu2020}). This geometric viewpoint is based on the time derivative of a monotonic functional of a fidelity-like measure (e.g., the Bures angle in the case of the Bures fidelity). We show that this approach can be potentially misleading, as it may suggest that different monotonic functionals of a given fidelity measure could yield different QSLs. In contrast, we prove that any monotone and differentiable reparametrization of a selected fidelity yields identical ML/MT bounds, thereby rendering the QSL invariant under such transformations within this geometric approach and providing a unified perspective that connects fidelity- and metric-induced geometric formulations. Consequently, we show that the QSLs depend solely on the chosen generalized fidelity measure, rather than on any intrinsic metric structure associated with its functional form (e.g., the Bures angle). This observation motivates the generalized framework developed in the present work, which establishes a unified inequality structure valid for a broad class of distinguishability measures.

Within this theoretical framework, we also analyze several concrete QSLs derived from distinguishability measures between quantum states, commonly referred to as generalized fidelities or alternative fidelities—that is, measures conceptually similar to the well-known Bures fidelity but differing in definition and properties. These generalized fidelities are monotonic and differentiable functions that quantify the distinguishability between two quantum states and typically decrease as the states become more distinguishable. In general, they satisfy some, but not necessarily all, of the properties of the Bures fidelity.

In addition, to explore the behavior of the QSLs analyzed in this work within a concrete physical setting, we consider the damped Jaynes–Cummings model, which describes the interaction between a two-level system and a single cavity mode that is, in turn, coupled to a reservoir composed of harmonic oscillators in the vacuum state.\par

This work is organized as follows. In Sec.~\ref{sec:buresfidelity}, we provide a brief overview of the Bures fidelity and its main properties. Sec.~\ref{sec:alternativefidelities} introduces a set of alternative fidelity measures that fulfill a minimal set of desirable properties for quantifying the similarity between quantum states. In Sec.~\ref{sec:analyticalresults}, we present our main analytical findings in the form of a proposition and a set of theorems. In Sec.~\ref{sec:QSLs-dampJCmodel}, we explore the behavior of the QSLs analyzed in this work within the physical setting provided by the damped Jaynes–Cummings model. Next, in Sec.~\ref{sec:analysis-discussion}, we analyze and discuss the analytical and numerical results obtained in this work. Finally, we introduce some concluding remarks in Sec.~\ref{sec:concludingremarks}.


\section{Bures fidelity}
\label{sec:buresfidelity}

The Bures fidelity $F_B$ (also known as the Uhlmann–Jozsa fidelity) is a widely used measure of the similarity between two density matrices. The Bures fidelity is defined as~\cite{bur69,Uhlm76,joz94}

\begin{equation*}
\label{UJFdef1}
F_B(\rho, \sigma) = \left[ \Tr\!\left(\sqrt{\sqrt{\rho}\, \sigma\, \sqrt{\rho}}\right) \right]^2,
\end{equation*}

where $\rho$ and $\sigma$ are arbitrary density matrices.

The Bures fidelity $F_B$ satisfies the four Jozsa axioms~\cite{joz94}:

\begin{enumerate}[label=(A\arabic*)]
\item \label{A1} \textit{Normalization}:
\begin{equation}
0 \leq F_B(\rho, \sigma) \leq 1\nonumber
\end{equation}
\item \label{A2} \textit{Symmetry}:
\begin{equation}
F_B(\rho, \sigma) = F_B(\sigma, \rho)\nonumber
\end{equation}
\item \label{A3} If $\rho = \ketbra{\psi}{\psi}$ represents a pure state, then
\[
F_B(\rho, \sigma) = \bra{\psi}\sigma\ket{\psi} = \Tr{\rho \sigma}\nonumber
\]
\item \label{A4} \textit{Unitary invariance}: For any arbitrary unitary process $\mathcal{U}$, $F_B(\rho, \sigma)$ is preserved i.e., 
\begin{equation}
F_B(\mathcal{U} \rho\, \mathcal{U}^\dag, \mathcal{U} \sigma \mathcal{U}^\dag) = F_B(\rho, \sigma)\nonumber
\end{equation}
\end{enumerate}

\noindent In addition to four Jozsa's axioms \ref{A1}--\ref{A4}, the Bures fidelity $F_B$ also satisfies the following properties ~\cite{joz94,nie00,Osan2013}:\par

\begin{enumerate}[label=(P\arabic*)]
\item \label{P1} {\em Identity of indiscernibles:}
\begin{equation}
F_B(\rho, \sigma) = 1 \:\:{\rm if\:\:and\:\:only\:\:if}\:\:\rho = \sigma\nonumber
\end{equation}
\item \label{P2}{\em Separate concavity:} For $p_1 , p_2 \geq 0$, $p_1+p_2=1$, and arbitrary density matrices $\rho_1$, $\rho_2$ and $\sigma$
\begin{equation}
F_B(p_1 \rho_1+p_2 \rho_2, \sigma) \geq p_1 F_B(\rho_1,\sigma) +p_2 F_B(\rho_2,\sigma)\nonumber
\end{equation}
By symmetry property~\ref{A2}, concavity in the second argument is also fulfilled.
\item \label{P3} {\em Multiplicativity under tensor product:} For arbitrary density matrices $\rho_1$, $\rho_2$, $\sigma_1$ and $\sigma_2$
\begin{equation}
F_B(\rho_1 \otimes \rho_2, \sigma_1 \otimes \sigma_2) = F_B(\rho_1,\sigma_1) F_B(\rho_2, \sigma_2)\nonumber
\end{equation}
\item \label{P4} {\em Monotonicity under quantum operations:} For a general quantum operation $\mathcal{E}$ described by a completely positive trace-preserving map (CPTP-map),
\begin{equation}
F_B(\mathcal{E}(\rho), \mathcal{E}(\sigma)) \geq F_B(\rho, \sigma)\nonumber
\end{equation}
\end{enumerate}

It is worth mentioning that $F_B$ can be physically interpreted as a generalized transition probability between two mixed states $\rho$ and $\sigma$. Some authors omit the square operation after performing the trace operation. In this case, the fidelity cannot be interpreted as a probability.\par

Generally speaking, the Bures fidelity can be difficult to compute. As a consequence, several alternative fidelity measures have been introduced in an effort to overcome this difficulty.


\section{Alternative fidelity measures}
\label{sec:alternativefidelities}

If one is willing to relax some of the properties of the Bures fidelity, a number of so-called generalized (or alternative) fidelities can be introduced. Below, we summarize several of these measures.

An upper bound of the Bures fidelity $F_B$, known as the \textit{super-fidelity} $\mathcal{F}_s$~\cite{Miszczak09,Mendo}, is defined as follows:
\begin{equation}
\label{eq:superfidelity}
\mathcal{F}_s(\rho,\sigma)
=\Tr{\rho\sigma}
+\sqrt{1-\Tr{\rho^{2}}}\sqrt{1-\Tr{\sigma^{2}}}.
\end{equation}

It is worth noting that, in the particular case of qubits, $\mathcal{F}_s$ coincides with the Bures fidelity $F_B$.

The super-fidelity $\mathcal{F}_s(\rho,\sigma)$ defined by Eq.~\eqref{eq:superfidelity} satisfies all of Jozsa’s axioms~\ref{A1}--\ref{A4}~\cite{Miszczak09,Mendo}. Additionally, $\mathcal{F}_s$ is super-multiplicative under tensor products~\cite{Miszczak09,Mendo}, i.e.,
\begin{eqnarray}
\mathcal{F}_s(\rho_{1}\otimes\rho_{2},\sigma_{1}\otimes\sigma_{2})
\geq
\mathcal{F}_s(\rho_{1},\sigma_{1})
\mathcal{F}_s(\rho_{2},\sigma_{2}).
\nonumber
\end{eqnarray}

The so-called operator fidelity is defined as follows~\cite{Wang2008}:
\begin{equation}
\label{eq:operfidel}
\mathcal{F}_o(\rho,\sigma)
= \frac{\Tr{\rho\sigma}}{\sqrt{\Tr{\rho^2}}\sqrt{\Tr{\sigma^2}}}.
\end{equation}

This quantity is often easier to compute than the Bures fidelity $F_B$. The operator fidelity $\mathcal{F}_o(\rho,\sigma)$ defined by Eq.~\eqref{eq:operfidel} satisfies Jozsa’s axioms~\ref{A1},~\ref{A2}, and~\ref{A4}, as well as properties~\ref{P1} and~\ref{P3}~\cite{Wang2008}.

If both the initial state $\rho$ and the state $\sigma$ represent pure states, it is straightforward to verify that the operator fidelity equals the Bures fidelity.

An alternative fidelity between quantum states was introduced in Ref.~\cite{Ektesabi2017}:
\begin{equation}
\label{eq:ektesabifidelity}
\mathcal{F}_a(\rho,\sigma)
=
\left(
1+
\sqrt{\frac{1-\Tr{\rho^2}}{\Tr{\rho^2}}}
\sqrt{\frac{1-\Tr{\sigma^2}}{\Tr{\sigma^2}}}
\right)
\Tr{\rho \sigma}.
\end{equation}

In particular, if the initial state $\rho_0$ represents a pure state, it is easy to see that this alternative fidelity equals the Bures fidelity.

The alternative fidelity $\mathcal{F}_a$ satisfies all of Jozsa’s axioms~\ref{A1}--\ref{A4} and vanishes when the two density matrices are orthogonal~\cite{Ektesabi2017}. Additionally, $\mathcal{F}_a$ is super-multiplicative under tensor products~\cite{Ektesabi2017}, i.e.,
\begin{eqnarray}
\mathcal{F}_a(\rho_{1}\otimes\rho_{2},\sigma_{1}\otimes\sigma_{2})
\geq
\mathcal{F}_a(\rho_{1},\sigma_{1})
\mathcal{F}_a(\rho_{2},\sigma_{2}).
\nonumber
\end{eqnarray}


\section{Analytical results}
\label{sec:analyticalresults}


\begin{proposition}
\label{proposition}
Let $D(x)$ be a bounded function defined on the domain $[0,1]$, assumed to be continuous, monotonically decreasing, and differentiable. Furthermore, suppose that its inverse function $D^{-1}(y)$ exists on the image of $D$ and is also differentiable.

Consider the following functional composition:

\begin{equation}
\label{eq:generalD}
\mathcal{L}_D(t) \doteq D\left(\mathcal{F}(\rho_0,\rho_t)\right) \equiv D(\mathcal{F}(t)),
\end{equation}

\noindent where $\mathcal{F}(\rho_0, \rho_t)\equiv \mathcal{F}(t)$ denotes a suitable measure of fidelity between an initial quantum state $\rho_0$ and the evolved state $\rho_t$ at time $t$. Under these assumptions, the following inequality holds:

\begin{equation}
\label{eq:newgenineq}
-\frac{d}{dt}D^{-1}\left(\mathcal{L}_D(t)\right) \leq \left| \dot{\mathcal{F}}(t) \right|.
\end{equation}

\flushright $\blacksquare$
\end{proposition}

\noindent \textit{Proof:}

Let us consider the composite function $\mathcal{L}_D(t) = D(\mathcal{F}(t))$, where the function $D(x)$ satisfies the properties stated in the proposition~\ref{proposition}. Therefore, it follows that the general fidelity $\mathcal{F}(t)$ can be expressed as $\mathcal{F}(t) = D^{-1}(\mathcal{L}_D(t))$. Now, we take the time derivative of Eq.~\eqref{eq:generalD}:

\begin{equation*}
\frac{d}{dt}\left( \mathcal{L}_D(\mathcal{F}(t)) \right) \leq \left| \frac{d}{dt} \left( \mathcal{L}_D(\mathcal{F}(t)) \right) \right| = \left| \frac{d \mathcal{L}_D}{d \mathcal{F}} \right| \cdot \left| \dot{\mathcal{F}}(t) \right|.
\end{equation*}

\noindent Hence, we obtain:

\begin{equation*}
\frac{d \mathcal{L}_D(t)}{dt} = \dot{\mathcal{L}}_D(t) = \frac{dD}{d\mathcal{F}} \cdot \frac{d\mathcal{F}(t)}{dt} \leq \left| \frac{d \mathcal{L}_D(t)}{dt} \right| = \left| \frac{dD}{d\mathcal{F}} \right| \cdot \left| \frac{d\mathcal{F}(t)}{dt} \right|.
\end{equation*}

Given that $D(x)$ is monotonically decreasing and differentiable, we have $dD/dx < 0$, implying that $|dD/dx| = -dD/dx$. Thus, the inequality simplifies to:

\begin{equation*}
\dot{\mathcal{L}}_D(t) \leq -\frac{dD}{d\mathcal{F}} \cdot \left| \frac{d\mathcal{F}(t)}{dt} \right|.
\end{equation*}


Next, we compute $dD/d\mathcal{F}$ by differentiating the inverse function (note that this approach uses the inverse function theorem in reverse form):

\begin{equation*}
\frac{dD}{d\mathcal{F}} = \left. \frac{1}{\frac{dD^{-1}(\mathcal{L}_D)}{d\mathcal{L}_D}} \right|_{\mathcal{L}_D = D(\mathcal{F})}.
\end{equation*}

\noindent Substituting this result into the previous inequality yields:

\begin{equation*}
\dot{\mathcal{L}}_D(t) \leq -\left( \frac{1}{\left. \frac{dD^{-1}(\mathcal{L}_D)}{d\mathcal{L}_D} \right|_{\mathcal{L}_D = D(\mathcal{F})}} \right) \cdot \left| \frac{d\mathcal{F}(t)}{dt} \right|.
\end{equation*}

\noindent This inequality can be rewritten as:

\begin{equation*}
- \frac{dD^{-1}(\mathcal{L}_D)}{d\mathcal{L}_D} \cdot \dot{\mathcal{L}}_D(t) \leq \left| \dot{\mathcal{F}}(t) \right|.
\end{equation*}

Finally, taking into account the chain rule for differentiating the composition $D^{-1}(\mathcal{L}_D(t))$, we arrive at:

\begin{equation*}
- \frac{d D^{-1}(\mathcal{L}_D(t))}{dt}  \leq \left| \dot{\mathcal{F}}(t) \right|,
\end{equation*}

\noindent which completes the proof of the proposition. \qed


\begin{theorem}
\label{theo:centraltheo}

Let $D(x)$ be a function that satisfies the properties stated in Proposition~\ref{proposition}.\par
Consider the following functional composition:
\begin{equation}
\mathcal{L}_D(t) \doteq D\left(\mathcal{F}(\rho_0,\rho_t)\right) \equiv D(\mathcal{F}(t)),
\end{equation}
where $\mathcal{F}(\rho_0, \rho_t)\equiv \mathcal{F}(t)$ denotes a suitable measure of fidelity between an initial quantum state $\rho_0$ and the evolved state $\rho_t$ at time $t$, such that $0\leq \mathcal{F}(t)\leq 1$, with $\mathcal{F}(0) = 1$.
Also, assume that there exists a function $B(t)$ which upper bounds the absolute value of the time derivative of $\mathcal{F}(t)$, i.e.,

\[
|\dot{\mathcal{F}}(t)| \leq B(t), \quad \forall t \geq 0.
\]

\noindent We define the time-averaged bound over the interval $0\leq t \leq\tau$ as

\begin{equation}
\label{eq:Btau}
\mathcal{B}(\tau) \doteq \frac{1}{\tau} \int_0^\tau B(t)\, dt.
\end{equation}

%

\noindent Then, the inequality

\begin{equation*}
- \frac{d D^{-1}(\mathcal{L}_D(t))}{dt}  \leq |\dot{\mathcal{F}}(t)|
\end{equation*}

\noindent leads to the following QSL:

\begin{equation}
\label{eq:genQSLineq}
\tau_{\mbox{\tiny QSL}} \doteq \frac{1 -\mathcal{F}(\tau)}{\mathcal{B}(\tau)}.
\end{equation}

\noindent The QSL expressed in Eq.~\eqref{eq:genQSLineq} holds universally, irrespective of the specific choice of the function $D(x)$, provided $D(x)$ satisfies the properties stated in Proposition~\ref{proposition}. \flushright $\blacksquare$\par

\end{theorem}

\noindent \textit{Proof:}\par

We start from Eq.~\eqref{eq:newgenineq}, i.e.,

\begin{equation*}
- \frac{d D^{-1}(\mathcal{L}_D(t))}{dt}  \leq |\dot{\mathcal{F}}(t)|,
\end{equation*}

as $|\dot{\mathcal{F}}(t)| \leq B(t)$ for $t \geq 0$, we can write

\begin{equation}
\label{eq:dotfidelitybound}
- \frac{d D^{-1}(\mathcal{L}_D(t))}{dt}  \leq B(t).
\end{equation}

Integrating Eq.~\eqref{eq:dotfidelitybound} over the time interval $0 \leq t \leq \tau$ gives the following QSL:

\begin{equation*}
\tau_{\mbox{\tiny QSL}} \doteq \frac{\left[ 1 - \mathcal{F}(\tau) \right]}{\mathcal{B}(\tau)},
\end{equation*}

\noindent where the quantity $\mathcal{B}(\tau)$ is given by Eq.~\eqref{eq:Btau}.

\noindent These last results complete the proof. \qed\par


\begin{corollary}[Immediate]
Under the hypotheses of Theorem~\ref{theo:centraltheo}, all strictly monotone differentiable post-processing functionals $g(\mathcal{F})$
generate the same QSL as using $\mathcal{F}$ itself.
Thus, within this geometric construction, no such reparametrization can produce a strictly tighter QSL.
\end{corollary}




To illustrate concrete examples derived from Proposition~\ref{proposition} and Theorem~\ref{theo:centraltheo}, we now introduce Theorems~\ref{theo:buresfidelity},~\ref{theo:superfidelity},~\ref{theo:operatorfidelity} and ~\ref{theo:Ektesabifidelity}, which follow directly from these results when a specific form of the similarity measure between quantum states is chosen.\par


\begin{theorem}
\label{theo:buresfidelity}
Let $D(x)$ be a function that satisfies the properties stated in Proposition~\ref{proposition}. Also, define the following functional composition:

\begin{equation*}
\mathcal{L}_D(t) \doteq D\big(F_B(\rho_0, \rho_t)\big) \equiv D(F_B(t)),
\end{equation*}

\noindent where $F_B$ denotes the Bures fidelity (c.f. Sec.~\ref{sec:buresfidelity}) which depends on the initial and final quantum states $\rho_0$ and $\rho_t$, respectively. Then, the following QSLs hold independently of the specific choice of the function $D(x)$, provided $D(x)$ satisfies the properties stated in Proposition~\ref{proposition}:\par

\noindent $\bullet$ ML-type QSLs:

$\circ$ \textit{Unitary (Hamiltonian) dynamics:}

\begin{equation}
\label{eq:genMLDeffHamiltdyn}
\tau_{\text{\tiny QSL}}^B \doteq \frac{\hbar}{2 E_\tau} \left[1 - F_B(\tau)\right],
\end{equation}

\noindent where $E_\tau$ is the time-averaged energy defined by:

\begin{equation}
\label{eq:deffneraverageenergy}
E_\tau \equiv \langle E \rangle_\tau = \frac{1}{\tau} \int_0^\tau \langle H_t \rangle \, \mathrm{d}t,
\end{equation}

\noindent and $\langle H_t \rangle \doteq \Tr{|H_t\rho(t)|}$.

$\circ$ \textit{Non-unitary evolution:}\par

\noindent Consider now the case of a general nonunitary evolution described by the time-local master equation:

\begin{equation*}
\dot{\rho}_t = L_t(\rho_t),
\end{equation*}

\noindent where $L_t$ is the dynamics generator~\cite{Lindblad1976,Gorini1976,breuer07}. The QSL in this case is:

\begin{equation}
\label{eq:genMLDeffLindbladevol}
\tau_{\text{\tiny QSL}}^B \doteq \max \left\{ \frac{1}{\mathcal{B}^{\mbox{\tiny (op)}}(\tau)}, \frac{1}{\mathcal{B}^{\mbox{\tiny (Tr)}}(\tau)} \right\} \left[1 - F_B(\tau)\right],
\end{equation}

\noindent where the quantities $\mathcal{B}^{\mbox{\tiny (op)}}(\tau)$ and $\mathcal{B}^{\mbox{\tiny (Tr)}}(\tau)$ are defined by:

\begin{align}
\label{eq:DeffnerMLBtauop}
\mathcal{B}^{\mbox{\tiny (op)}}(\tau) &= \frac{1}{\tau} \int_0^\tau \; \left\| L_t(\rho_t) \right\|_{\mbox{\tiny op}}  \mathrm{d}t, \\
\mathcal{B}^{\mbox{\tiny (Tr)}}(\tau) &= \frac{1}{\tau} \int_0^\tau \; \left\| L_t(\rho_t) \right\|_{\mbox{\tiny tr}}\mathrm{d}t.
\label{eq:DeffnerMLBtauTr}
\end{align}

%
\noindent $\bullet$ MT-type QSL

\noindent The QSL in this case is:\par

\begin{equation}
\label{eq:MTboundDeffLindbladevol}
\tau_{\mbox{\tiny QSL}}^B \doteq \frac{1}{\mathcal{B}^{(\mbox{\tiny HS})}(\tau)} \left[1 - F_B(\tau)\right],
\end{equation}

\noindent where

\begin{equation}
\label{eq:deffnerMTBtauHS}
\mathcal{B}^{(\mbox{\tiny HS})}(\tau) = \frac{1}{\tau} \int_0^\tau \mathrm{d}t \; \left\| L_t(\rho_t) \right\|_{\mbox{\tiny HS}},
\end{equation}

\noindent and the Hilbert–Schmidt norm is defined, for a given operator $A$, as

\[
\left\| A \right\|_{\mbox{\tiny HS}} = \sqrt{\Tr{A^\dagger A}} = \sqrt{\sum_i \sigma_i^2},
\]

\noindent where $\{\sigma_i\}$ are the singular values of $A$~\cite{bha97}. \flushright$\blacksquare$\par

\end{theorem}


\noindent \textit{Proof:}

\noindent $\bullet$ \textit{ML-type QSLs}

\vspace{0.25truecm}
$\circ$ \textit{Hamiltonian Dynamics:}

\noindent Consider a quantum system evolving under a time-dependent Hamiltonian $H_t$. The time evolution of the system’s density operator $\rho_t$ is governed by the von Neumann's equation:

\begin{equation}
\label{eq:vonNeumanneq}
\dot\rho_t = \frac{1}{i\hbar} \left[H_t, \rho_t\right].
\end{equation}

Assuming the system starts in a pure state $\rho_0 = \ketbra{\psi_0}{\psi_0}$, Eq.~\eqref{eq:newgenineq} in proposition \eqref{proposition} becomes

\begin{equation}
\label{newgenineqpure}
-\frac{dD^{-1}(\mathcal{L}_D(t))}{dt} \leq \left| \bra{\psi_0} \dot{\rho}_t \ket{\psi_0} \right|.
\end{equation}

\noindent Substituting Eq.~\eqref{eq:vonNeumanneq} into Eq.~\eqref{newgenineqpure}, we obtain:

\begin{equation*}
-\frac{dD^{-1}(\mathcal{L}_D(t))}{dt} \leq \frac{1}{\hbar} \left| \bra{\psi_0} [H_t, \rho_t] \ket{\psi_0} \right|.
\end{equation*}

\noindent Applying the triangle inequality to the right-hand side, we arrive at:

\begin{equation*}
-\frac{dD^{-1}(\mathcal{L}_D(t))}{dt} \leq \frac{1}{\hbar} \left( \left| \Tr{H_t \rho_t \rho_0} \right| + \left| \Tr{\rho_t H_t \rho_0} \right| \right).
\end{equation*}

To estimate an upper bound for the terms on the right-hand side, we use the von Neumann's trace inequality ~\cite{vonNeumann1937,Mirsky1975,Grigorieff1991,Deffner2013}. Taking $A = (H_t \rho_t)^\dagger = \rho_t H_t$ and applying von Neumann's trace inequality ~\cite{vonNeumann1937,Mirsky1975,Grigorieff1991,Deffner2013}, we find:

\begin{equation*}
-\frac{dD^{-1}(\mathcal{L}_D(t))}{dt} \leq \frac{2}{\hbar} \sum_i \sigma_i p_i = \frac{2}{\hbar} \sigma_1,
\end{equation*}

\noindent where $\sigma_i$ are the singular values of $H_t \rho_t$ (with $\sigma_1 \geq \sigma_2 \geq \cdots \geq \sigma_n$), and $p_i = \delta_{i,1}$ follows from the pure initial state $\rho_0$. For Hermitian operators, the operator norm equals the largest singular value, i.e., $||A||_{\mbox{\tiny op}} = \sigma_1$, while the trace norm equals the sum of all singular values, i.e., $||A||_{\mbox{\tiny Tr}} = \sum_i \sigma_i$~\cite{bha97}. Hence,

\begin{equation}
\label{eq10}
-\frac{dD^{-1}(\mathcal{L}_D(t))}{dt} \leq \frac{2}{\hbar} \| H_t \rho_t \|_{\mbox{\tiny op}} \leq \frac{2}{\hbar} \| H_t \rho_t \|_{\mbox{\tiny Tr}}.
\end{equation}

If the instantaneous eigenvalues of $H_t$ are positive (which can always be ensured by an appropriate choice of the energy zero~\cite{Margolus1998}), the trace norm becomes:

\begin{equation*}
\| H_t \rho_t \|_{\mbox{\tiny Tr}} = \Tr{\left| H_t \rho_t \right|} = \langle H_t \rangle.
\end{equation*}

\noindent Thus, equation \eqref{eq10} can be written as:

\begin{equation}
\label{eq10bis}
-\frac{dD^{-1}(\mathcal{L}_D(t))}{dt} \leq \langle H_t \rangle.
\end{equation}

\noindent Integrating Eq.~\eqref{eq10bis} over the time interval $0 \leq t \leq \tau$ yields

\begin{equation*}
\tau_{\mbox{\tiny QSL}}^B \doteq \frac{\hbar}{2E_\tau} \left[ 1 - F_B(\tau) \right],
\end{equation*}

\noindent where $E_\tau$ is given by Eq.~\eqref{eq:deffneraverageenergy}.\par

$\circ$ \textit{Non-Unitary Evolution:}

\noindent Now consider an arbitrary time-dependent nonunitary evolution governed by

\begin{equation}
\label{eq13}
\dot\rho_t = L_t(\rho_t),
\end{equation}

\noindent where $L_t$ is the dynamics generator~\cite{Lindblad1976,Gorini1976,breuer07}. These generators are trace-class superoperators in a complex Banach space and may be non-Hermitian. However, for symmetric norms such as the Schatten $p$-norm, $||L_t||_p = \left( \sum_i \lambda_i^p \right)^{1/p}$~\cite{rem2}, the equality $||L_t^\dagger|| = ||L_t||$ holds, ensuring the validity of previous equations and bounds~\cite{sim05,Deffner2013}.

Substituting Eq.~\eqref{eq13} into Eq.~\eqref{newgenineqpure} yields:

\begin{equation*}
-\frac{dD^{-1}(\mathcal{L}_D(t))}{dt} \leq \left| \bra{\psi_0} L_t(\rho_t) \ket{\psi_0} \right| = \left| \Tr{L_t(\rho_t) \rho_0} \right|.
\end{equation*}

\noindent Applying the von Neumann's trace inequality~\cite{vonNeumann1937,Mirsky1975,Grigorieff1991,Deffner2013} again, we obtain

\begin{equation}
\label{eq16}
-\frac{dD^{-1}(\mathcal{L}_D(t))}{dt} \leq \sum_i \lambda_i p_i = \lambda_1,
\end{equation}

\noindent where ${\lambda_i}$ (with $\lambda_1 \geq \lambda_2 \geq \cdots \geq \lambda_n$) are the singular values of $L_t(\rho_t)$. Using the norm hierarchy, we further bound Eq.~\eqref{eq16}:

\begin{equation}
\label{eq:MLineqDeffner}
-\frac{dD^{-1}(\mathcal{L}_D(t))}{dt} \leq \| L_t(\rho_t) \|_{\mbox{\tiny op}} \leq \| L_t(\rho_t) \|_{\mbox{\tiny Tr}}.
\end{equation}

In analogy with Ref.~\cite{Deffner2013}, integrating Eq.~\eqref{eq:MLineqDeffner} over the time interval $0\leq t \leq \tau$ allows us to obtain a ML-type QSL for open quantum systems as follows

\begin{equation*}
\tau_{\mbox{\tiny QSL}}^B \doteq \max \left\{ \frac{1}{\mathcal{B}_\tau^{\mbox{\tiny (op)}}}, \frac{1}{\mathcal{B}_\tau^{\mbox{\tiny (Tr)}}} \right\} \left[1 - F_B(\tau) \right],
\end{equation*}

\noindent where $\mathcal{B}^{\mbox{\tiny (op)}}(\tau)$ and $\mathcal{B}^{\mbox{\tiny (Tr)}}(\tau)$ are given by Eqs. \eqref{eq:DeffnerMLBtauop} and \eqref{eq:DeffnerMLBtauTr}.

\noindent $\bullet$ \textit{MT-type QSL}

To derive a MT-type QSL in the case of arbitrary positive open-system dynamics, we follow refs.~\cite{delCampo2013} and~\cite{Deffner2013}.

Substituting Eq.~\eqref{eq13} into Eq.~\eqref{newgenineqpure} yields:

\begin{equation*}
-\frac{dD^{-1}(\mathcal{L}_D(t))}{dt} \leq \left| \Tr{L_t(\rho_t) \rho_0} \right|.
\end{equation*}

\noindent Applying the Cauchy–Schwarz inequality for operators~\cite{delCampo2013,Deffner2013} we obtain:

\begin{equation*}
-\frac{dD^{-1}(\mathcal{L}_D(t))}{dt} \leq \sqrt{ \Tr{L_t(\rho_t) L_t(\rho_t)^\dagger} \Tr{\rho_0^2} }.
\end{equation*}

For a pure initial state $\rho_0$, $\Tr{\rho_0^2} = 1$, hence:

\begin{equation}
\label{eq21}
-\frac{dD^{-1}(\mathcal{L}_D(t))}{dt} \leq \sqrt{ \Tr{L_t(\rho_t) L_t(\rho_t)^\dagger} } = \| L_t(\rho_t) \|_{\mbox{\tiny HS}},
\end{equation}
where the Hilbert–Schmidt norm is defined as $\| A \|_{\mbox{\tiny HS}} = \sqrt{\Tr{A^\dagger A}} = \sqrt{ \sum_i \sigma_i^2 }$~\cite{bha97}. Integrating Eq.~\eqref{eq21} over the time interval $0 \leq t \leq \tau$ gives the following QSL:

\begin{equation*}
\tau_{\mbox{\tiny QSL}}^B \doteq \frac{\left[ 1 - F_B(\tau) \right]}{\mathcal{B}^{\mbox{\tiny (HS)}}(\tau)},
\end{equation*}

\noindent where the quantity $\mathcal{B}^{\mbox{\tiny (HS)}}(\tau)$ is given by Eq.~\eqref{eq:deffnerMTBtauHS}.

\noindent These last results complete the proof. \qed\par


\begin{theorem}
\label{theo:superfidelity}

Let $D(x)$ be a function that satisfies the properties stated in Proposition~\ref{proposition}. Also, define the following functional composition:

\begin{equation*}
\mathcal{L}_D(t) \doteq D\big(\mathcal{F}_s(\rho_0, \rho_t)\big) \equiv D(\mathcal{F}_s(t)),
\end{equation*}

\noindent where $\mathcal{F}_s(t)$ denotes the super-fidelity given by

\[
\mathcal{F}_s(t) =\Tr{\rho_0 \rho_t} + \sqrt{1-\Tr{\rho_0^{2}}}\sqrt{1-\Tr{\rho_t^{2}}}.
\]

\noindent (cf.~Eq.~\eqref{eq:superfidelity}), Sec.~\ref{sec:alternativefidelities}), which depends on the initial and final quantum states $\rho_0$ and $\rho_t$, respectively. Consider the case of a general nonunitary evolution described by the time-local master equation:

\begin{equation*}
\dot{\rho}_t = L_t(\rho_t),
\end{equation*}

\noindent where $L_t$  is the dynamics generator~\cite{Lindblad1976,Gorini1976,breuer07}. Then, the following QSLs hold independently of the specific choice of the function $D(x)$, provided $D(x)$ satisfies the properties stated in Proposition~\ref{proposition}:

\noindent $\bullet$ \textit{ML-type QSLs}:

\noindent The QSL in this case is:

\begin{equation}
\label{eq:MLboundsuperfidelity}
\tau_{\text{\tiny QSL}}^s \doteq \max \left\{ \frac{1}{\mathcal{B}^{\mbox{\tiny (op)}}_s(\tau)}, \frac{1}{\mathcal{B}^{\mbox{\tiny (Tr)}}_s(\tau)}\right\} \left[1-\mathcal{F}_s(\tau)\right],
\end{equation}

\noindent where the quantities $\mathcal{B}^{\mbox{\tiny (op)}}_s(\tau)$ and $\mathcal{B}^{\mbox{\tiny (Tr)}}_s(\tau)$ are given by

\begin{align}
\mathcal{B}^{\mbox{\tiny (op)}}_s(\tau) &\doteq \frac{1}{\tau} \int_0^\tau \left|\left\| L_t(\rho_t)\right|\right\|_{\mbox{\tiny op}}\cdot\left(1+\sqrt{\frac{1 - \Tr{\rho_0^2}}{1 - \Tr{\rho_t^2}}}\right)\, dt\label{eq:superfidelityMLBtauop},\\
\mathcal{B}^{\mbox{\tiny (Tr)}}_s(\tau) &\doteq \frac{1}{\tau} \int_0^\tau \left|\left| L_t(\rho_t)\right|\right|_{\mbox{\tiny Tr}}\cdot\left(1+\sqrt{\frac{1 - \Tr{\rho_0^2}}{1 - \Tr{\rho_t^2}}}\right)\, dt.\label{eq:superfidelityMLBtauTr}
\end{align}

\noindent $\bullet$ \textit{MT-type QSL}:

\noindent The QSL in this case is:

\begin{align}
\label{eq:MTboundsuperfidelity}
\tau_{\mbox{\tiny QSL}}^s = \frac{\left[1-\mathcal{F}_s(\tau) \right]}{\mathcal{B}^{\mbox{\tiny (HS)}}_s(\tau)},
\end{align}

\noindent where

\begin{align}
\mathcal{B}^{\mbox{\tiny (HS)}}_s(\tau) = \frac{1}{\tau} \int_0^\tau dt\, \|L_t(\rho_t)\|_{\mbox{\tiny HS}} \cdot \left(1 + \sqrt{\frac{1 - \Tr{\rho_0^2}}{1 - \Tr{\rho_t^2}}} \right). \label{eq:superfidelityMTBtauHS}
\end{align}

\flushright$\blacksquare$\par

\end{theorem}


\noindent \textit{Proof:}\par

We start from Eq.~\eqref{eq:newgenineq}, i.e.,

\begin{equation}
\label{eq:ineqsuperfidelity1}
-\frac{d}{dt}D^{-1}\left(\mathcal{L}_D(t)\right) \leq \left| \dot{\mathcal{F}}_s(t) \right|,
\end{equation}

\noindent and, bearing in mind Ref.~\cite{Wu2020}, we notice that the absolute value of the time derivative of the super-fidelity $\mathcal{F}_s(\rho_0,\rho_t)$ is given by

\begin{align}
\left|\dot{\mathcal{F}}_s(t)\right| &= \left| \Tr{\rho_0 \dot{\rho}_t} - \sqrt{\frac{1 - \Tr{\rho_0^2}}{1 - \Tr{\rho_t^2}}} \, \Tr{\rho_t \dot{\rho}_t} \right| \nonumber \\
&\leqslant \left| \Tr{\rho_0 \dot{\rho}_t} \right| + \sqrt{\frac{1 - \Tr{\rho_0^2}}{1 - \Tr{\rho_t^2}}} \left| \Tr{\rho_t \dot{\rho}_t} \right|.
\label{eq:moddotFs}
\end{align}

For general non-unitary dynamics, the evolution of the density operator is governed by a time-local master equation of the form $\dot{\rho}_t = L_t(\rho_t)$. Substituting \eqref{eq:moddotFs} into Eq.~\eqref{eq:ineqsuperfidelity1}, we obtain:

\begin{align}
- \frac{dD^{-1}(\mathcal{L}_D(t))}{dt} \leqslant& \left| \Tr{\rho_0 L_t(\rho_t)} \right|\nonumber\\
 + &\sqrt{\frac{1 - \Tr{\rho_0^2}}{1 - \Tr{\rho_t^2}}} \left| \Tr{\rho_t L_t(\rho_t)} \right|. \label{eq8}
\end{align}

\noindent $\bullet$ \textit{ML-type QSLs}\par

Applying von Neumann's trace inequality~\cite{vonNeumann1937,Mirsky1975,Grigorieff1991,Deffner2013} to the first term on the right-hand side of Eq.~\eqref{eq8}, we obtain

\begin{align*}
\left| \Tr{\rho_0 L_t(\rho_t)} \right| \leqslant \sum_i p_i \lambda_i,
\end{align*}

\noindent where $\{p_i\}$ (with $p_1 \geq p_2 \geq \cdots \geq p_n$) and ${\lambda_i}$ (with $\lambda_1 \geq \lambda_2 \geq \cdots \geq \lambda_n$) are the singular values of $\rho_0$ and $L_t(\rho_t)$, respectively. Similarly, for the second term:

\begin{align*}
\left| \Tr{\rho_t L_t(\rho_t)} \right| \leqslant \sum_i \epsilon_i \lambda_i,
\end{align*}

\noindent where $\{\epsilon_i\}$ (with $\epsilon_1 \geq \epsilon_2 \geq \cdots \geq \epsilon_n$) are the singular values of $\rho_t$.

Since $p_i \leq 1$ and $\epsilon_i \leq 1$, it follows that

\[
\sum_i p_i \lambda_i \leq \lambda_1 \leq \sum_i \lambda_i, \qquad \sum_i \epsilon_i \lambda_i \leq \lambda_1 \leq \sum_i \lambda_i,
\]

\noindent where $\lambda_1$ is the largest singular value of $L_t(\rho_t)$, which corresponds to the operator norm $\|L_t(\rho_t)\|_{\mbox{\tiny op}}$, while the sum $\sum_i \lambda_i$ gives the trace norm $\|L_t(\rho_t)\|_{\mbox{\tiny Tr}}$. Thus, we can write

\begin{align}
\label{eq:MLineqsuperfidelity}
- \frac{dD^{-1}(\mathcal{L}_D(t))}{dt} \leqslant \left|\left| L_t(\rho_t)\right|\right|_{\mbox{\tiny op},\mbox{\tiny Tr}}\left(1+\sqrt{\frac{1 - \Tr{\rho_0^2}}{1 - \Tr{\rho_t^2}}}\right)
\end{align}

In analogy with Ref.~\cite{Deffner2013}, integrating Eq.~\eqref{eq:MLineqsuperfidelity} over the time interval $0\leq t \leq \tau$ allows us to obtain a ML-type QSL for open quantum systems as follows

\begin{align*}
\tau_{\mbox{\tiny QSL}}^s = \max \left\{ \frac{1}{\mathcal{B}^{\mbox{\tiny (op)}}_s(\tau)}, \frac{1}{\mathcal{B}^{\mbox{\tiny (Tr)}}_s(\tau)} \right\} \left[ 1 - \mathcal{F}_s(\tau) \right], 
\end{align*}


\noindent where the time-averaged quantities $\mathcal{B}^{\mbox{\tiny (op)}}_s(\tau)$ and $\mathcal{B}^{\mbox{\tiny (Tr)}}_s(\tau)$ are given by Eqs.~\eqref{eq:superfidelityMLBtauop} and \eqref{eq:superfidelityMLBtauTr}.

\noindent $\bullet$ \textit{MT-type QSL}\par

Using the Cauchy–Schwarz inequality for operators~\cite{delCampo2013,Deffner2013}, we can bound Eq.~\eqref{eq8} in the following way:

\begin{align}
- \frac{dD^{-1}(\mathcal{L}_D(t))}{dt} 
&\leq \sqrt{\Tr{\rho_0^2}} \sqrt{\Tr{L_t^\dag(\rho_t) L_t(\rho_t)}} \nonumber \\
+& \sqrt{\frac{1 - \Tr{\rho_0^2}}{1 - \Tr{\rho_t^2}}} \sqrt{\Tr{\rho_t^2}} \sqrt{\Tr{L_t^\dag(\rho_t) L_t(\rho_t)}} \nonumber \\
&\leq \left( 1 + \sqrt{\frac{1 - \Tr{\rho_0^2}}{1 - \Tr{\rho_t^2}}} \right) \|L_t(\rho_t)\|_{\mbox{\tiny HS}},
\label{eq:MTineqsuperfidelity}
\end{align}

\noindent where the last step uses the fact that $\Tr{\rho^2} \leq 1$ for any density operator, and $\|L_t(\rho_t)\|_{\mbox{\tiny HS}} = \sqrt{\sum_i \lambda_i^2}$ denotes the Hilbert–Schmidt norm of $L_t(\rho_t)$.

Integrating Eq.~\eqref{eq:MTineqsuperfidelity} over the time interval $0 \leq t \leq \tau$ gives the following  MT-type QSL for non-unitary evolution:


\begin{align*}
\tau_{\mbox{\tiny QSL}}^s \doteq \frac{\left[1-\mathcal{F}_s(\tau) \right]}{\mathcal{B}^{\mbox{\tiny (HS)}}_s(\tau)},
\end{align*}

\noindent where $\mathcal{B}^{\mbox{\tiny (HS)}}_s(\tau)$ is given by Eq.~\eqref{eq:superfidelityMTBtauHS}

\noindent These last results complete the proof. \qed\par


\begin{theorem}
\label{theo:operatorfidelity}
Let $D(x)$ be a function that satisfies the properties stated in Proposition~\ref{proposition}. Also, define the following functional composition:

\begin{equation}
\mathcal{L}_D(t) \doteq D\big(\mathcal{F}_o(\rho_0, \rho_t)\big) \equiv D(\mathcal{F}_o(t)),
\end{equation}

\noindent where $\mathcal{F}_o(t)$ denotes the operator fidelity  given by


\begin{equation*}
    \mathcal{F}_o(t) = \frac{\big|\Tr{\rho_0\rho_t}\big|}{\sqrt{\Tr{\rho_0^2}}\sqrt{\Tr{\rho_t^2}}},
\end{equation*} 

\noindent (cf.~Eq.~\eqref{eq:operfidel}), Sec.~\ref{sec:alternativefidelities}, which depends on the initial and final quantum states $\rho_0$ and $\rho_t$, respectively. Consider the case of a general nonunitary evolution described by the time-local master equation:

\begin{equation*}
\dot{\rho}_t = L_t(\rho_t),
\end{equation*}

\noindent where $L_t$  is the dynamics generator~\cite{Lindblad1976,Gorini1976,breuer07}. Then, the following MT-type QSL holds independently of the specific choice of the function $D(x)$, provided $D(x)$ satisfies the properties stated in Proposition~\ref{proposition}:


\begin{equation}
\label{eq:MTQSLoperfidel}
\tau_{\text{\tiny QSL}}^o \doteq \frac{1 - \mathcal{F}_o(\tau)}{\mathcal{B}_o(\tau)},
\end{equation}

\noindent where the quantity $\mathcal{B}_o(\tau)$ is given by the following equation:

\begin{equation}
 \label{eq:operatorfidelityMTBtau}
\mathcal{B}_o(\tau)\doteq\frac{2}{\tau}\int_{0}^{\tau} \sqrt{\frac{\Tr{\dot{\rho}_t^2}}{\Tr{\rho_t^2}}}\, dt
\end{equation}

\end{theorem}

\noindent \textit{Proof:}

We start from Eq.~\eqref{eq:newgenineq}, i.e.,

\begin{equation*}
-\frac{d}{dt}D^{-1}\left(\mathcal{L}_D(t)\right) \leq \left| \dot{\mathcal{F}_o}(t) \right|,
\end{equation*}

\noindent and, bearing in mind Ref.~\cite{Sun2015}, we notice that the absolute value of the time derivative of the operator fidelity $\mathcal{F}_o(t)$ is given by

\begin{equation*}
\left| \dot{\mathcal{F}_o}(t) \right| = \left| \frac{\Tr{\rho_0\dot{\rho}_t}}{\sqrt{\Tr{\rho_0^2}}\sqrt{\Tr{\rho_t^2}}} - \frac{\Tr{\rho_0 \rho_t}\Tr{\rho_t\dot{\rho}_t}}{\sqrt{\Tr{\rho_0^2}}\left[\Tr{\rho_t^2}\right]^{3/2}}\right|
\end{equation*}

Applying the triangle inequality to the right-hand side, we obtain:

\begin{equation}
\label{eq:ineqoperatorfidelity1}
\left| \dot{\mathcal{F}_o}(t) \right| \leq \frac{\left| \Tr{\rho_0\dot{\rho}_t}\right|}{\sqrt{\Tr{\rho_o^2}}\sqrt{\Tr{\rho_t^2}}} + \frac{\left|\Tr{\rho_0 \rho_t}\Tr{\rho_t\dot{\rho}_t}\right|}{\sqrt{\Tr{\rho_0^2}}\left[\Tr{\rho_t^2}\right]^{3/2}}
\end{equation}

\noindent Now, using the Cauchy–Schwarz inequality for operators~\cite{delCampo2013,Deffner2013} on the right side of Eq.~\eqref{eq:ineqoperatorfidelity1}, we can write:

\begin{equation}
\label{eq:ineqoperatorfidelity2}
-\frac{d}{dt}D^{-1}\left(\mathcal{L}_D(t)\right) \leq 2 \sqrt{\frac{\Tr{\dot{\rho}_t^2}}{\Tr{\rho_t^2}}}
\end{equation}

\noindent Integrating Eq.~\eqref{eq:ineqoperatorfidelity2} over the time interval $0\leq t \leq \tau$ allows us to obtain a QSL as follows:

\begin{equation}
\tau_{\text{\tiny QSL}}^o \doteq \frac{1 - \mathcal{F}_o(\tau)}{\mathcal{B}_o(\tau)},
\end{equation}

\noindent where the quantity $\mathcal{B}_o(\tau)$ is given by \eqref{eq:operatorfidelityMTBtau}.

\noindent These last results complete the proof. \qed\par


\begin{theorem}
\label{theo:Ektesabifidelity}
Let $D(x)$ be a function that satisfies the properties stated in Proposition~\ref{proposition}. Also, define the following functional composition:

\begin{equation*}
\mathcal{L}_D(t) \doteq D\big(\mathcal{F}_a(\rho_0, \rho_t)\big) \equiv D(\mathcal{F}_a(t)),
\end{equation*}

\noindent where $\mathcal{F}_a(t)$ denotes the alternative fidelity  given by

\[
\mathcal{F}_a(t)=\biggr(1+\sqrt{\frac{1-\Tr{\rho_0^2}}{\Tr{\rho_0^2}}}\sqrt{\frac{1-\Tr{\rho_t^2}}{\Tr{\rho_t^2}}}\biggr)\Tr{\rho_0 \rho_t},
\]

\noindent (cf.~Eq.~\eqref{eq:ektesabifidelity}), Sec.~\ref{sec:alternativefidelities}), which depends on the initial and final quantum states $\rho_0$ and $\rho_t$, respectively. Consider the case of a general nonunitary evolution described by the time-local master equation:

\begin{equation*}
\dot{\rho}_t = L_t(\rho_t),
\end{equation*}

\noindent where $L_t$ is the dynamics generator~\cite{Lindblad1976,Gorini1976,breuer07}. Then, the following MT-type QSL holds independently of the specific choice of the function $D(x)$, provided $D(x)$ satisfies the properties stated in Proposition~\ref{proposition}:


\begin{equation*}
\label{eq:MTQSLektesabifidel}
\tau_{\text{\tiny QSL}}^a \doteq \frac{1 - \mathcal{F}_a(\tau)}{\mathcal{B}_a(\tau)},
\end{equation*}

\noindent where the quantity $\mathcal{B}_a(\tau)$ is given by the following equation:

\begin{widetext}
 \begin{eqnarray}
 \label{eq:ektesabifidelityMTBtau}
 \begin{split}
 \mathcal{B}_a(\tau)\doteq\frac{1}{\tau}\int_{0}^{\tau}\biggr(\sqrt{\frac{1-\Tr{\rho_{0}^2}}{\Tr{\rho_{0}^2}}}\sqrt{\frac{\Tr{\rho_{t}^2}}{1-\Tr{\rho_{t}^2}}}\biggr|\frac{\Tr{\dot{\rho_{t}}\rho_{t}}\Tr{\rho_{0}\rho_{t}}}{(\Tr{\rho^2_{t}})^2}\biggr|
+
\sqrt{\Tr{\rho_{0}^2}\Tr{\dot{\rho_{t}}^2}}+\\
\sqrt{\frac{1-\Tr{\rho_{t}^2}}{\Tr{\rho_{t}^2}}}\sqrt{1-\Tr{\rho_{0}^2}}\sqrt{\Tr{\dot{\rho_{t}}^2}}\biggr)dt.
\end{split}
\end{eqnarray}
\end{widetext}

\end{theorem}

\noindent \textit{Proof:}\par

We start from Eq.~\eqref{eq:newgenineq}, i.e.,

\begin{equation*}
-\frac{d}{dt}D^{-1}\left(\mathcal{L}_D(t)\right) \leq \left| \dot{\mathcal{F}_a}(t) \right|,
\end{equation*}

\noindent and, bearing in mind Ref.~\cite{Ektesabi2017}, we notice that the absolute value of the time derivative of the alternative fidelity $\mathcal{F}_a(t)$ is given by

\begin{widetext}
\begin{eqnarray*}
\begin{split}
\left|\frac{d\mathcal{F}_a(t)}{dt}\right|
=\biggr|\frac{-\Tr{\dot{\rho_{t}}\rho_{t}}}{(\Tr{\rho_{t}})^2}
\sqrt{\frac{1-\Tr{\rho_{0}^2}}{\Tr{\rho_{0}^2}}} \sqrt{\frac{\Tr{\rho_{t}^2}}{1-\Tr{\rho_{t}^2}}}\Tr{\rho_{0}\rho_{t}}
+
\left(1+\sqrt{\frac{1-\Tr{\rho_{0}^2}}{\Tr{\rho_{0}^2}}}
\sqrt{\frac{1-\Tr{\rho_{t}^2}}{\Tr{\rho_{t}^2}}}\right)\Tr{\rho_{0}\dot{\rho_{t}}}\biggr|.
\end{split}
\end{eqnarray*}
\end{widetext}

\noindent Applying the triangle inequality we can write

\begin{widetext}
\begin{eqnarray}
\begin{split}
\left|\frac{d\mathcal{F}_a(t)}{dt}\right|
\leq
\sqrt{\frac{1-\Tr{\rho_{0}^2}}{\Tr{\rho_{0}^2}}}\sqrt{\frac{\Tr{\rho_{t}^2}}{1-\Tr{\rho_{t}^2}}}\left|\frac{\Tr{\dot{\rho_{t}}\rho_{t}}\Tr{\rho_{0}\rho_{t}}}{(\Tr{\rho_{t}})^2}\right|
+
\left(1+\sqrt{\frac{1-\Tr{\rho_{0}^2}}{\Tr{\rho_{0}^2}}}
\sqrt{\frac{1-\Tr{\rho_{t}^2}}{\Tr{\rho_{t}^2}}}\right)\left|\Tr{\rho_{0}\dot{\rho_{t}}}\right|.\label{eq:ineqEktesabifidelity1}
\end{split}
\end{eqnarray}
\end{widetext}

Thus, by using Cauchy-Schwarz inequality~\cite{delCampo2013,Deffner2013} on the second term of Eq.~\eqref{eq:ineqEktesabifidelity1}, we obtain:\par

\begin{widetext}
\begin{eqnarray}
\begin{split}
-\frac{d}{dt}D^{-1}\left(\mathcal{L}_D(t)\right) \leq\sqrt{\frac{1-\Tr{\rho_{0}^2}}{\Tr{\rho_{0}^2}}}\sqrt{\frac{\Tr{\rho_{t}^2}}{1-\Tr{\rho_{t}^2}}}\biggr|\frac{\Tr{\dot{\rho_{t}}\rho_{t}}\Tr{\rho_{0}\rho_{t}}}{(\Tr{\rho_{t}})^2}\biggr|
+
\sqrt{\Tr{\rho_{0}^2}\Tr{\dot{\rho_{t}}^2}}+\\
\sqrt{\frac{1-\Tr{\rho_{t}^2}}{\Tr{\rho_{t}^2}}}\sqrt{1-\Tr{\rho_{0}^2}}\sqrt{\Tr{\dot{\rho_{t}}^2}}.\label{eq:ineqEktesabifidelity2}
\end{split}
\end{eqnarray}
\end{widetext}

\noindent Integrating Eq.~\eqref{eq:ineqEktesabifidelity2} over the time interval $0\leq t \leq \tau$ allows us to obtain a QSL as follows

\begin{equation*}
\tau_{\mbox{\tiny QSL}}^a \doteq \frac{1 - \mathcal{F}_a(\tau)}{\mathcal{B}_a(\tau)},
\end{equation*}
 
\noindent where the quantity $\mathcal{B}_a(\tau)$ is given by Eq.~\eqref{eq:ektesabifidelityMTBtau}.

\noindent These last results complete the proof. \qed\par


\section{Quantum speed limits in a physical model}
\label{sec:QSLs-dampJCmodel}
To analyze the behavior of the QSLs in a concrete physical setting, we shall consider the damped Jaynes–Cummings model, which admits an exact solution. This model describes the interaction between a two-level system and a single cavity mode, which in turn is coupled to a reservoir composed of harmonic oscillators in the vacuum state. When we restrict the dynamics to the single-excitation subspace of the system–cavity composite, the cavity mode can be eliminated by means of an effective spectral density of the form~\cite{breuer07, Breuer1999}

\begin{equation*}
\label{eq:JCmodel}
J(\omega)=\frac{1}{2\pi}\frac{\gamma_0\lambda^2}
{(\omega_{0}-\omega)^2+\lambda^2},
\end{equation*}

\noindent where $\omega_{0}$ denotes the transition frequency of the two-level system. The parameter $\lambda$ characterizes the spectral width of the coupling and is related to the reservoir correlation time $\tau_{R}$ through $\tau_{R}=\lambda^{-1}$. The characteristic time scale over which the state of the system evolves is given by $\tau_{S}=\gamma_0^{-1}$. In this model, the nonunitary generator of the reduced dynamics of the system is given by
  
\begin{equation}
\begin{split}
L_t(\rho_t) =\gamma_t\left(\sigma_{-}\rho_t\sigma_{+} -\frac{1}{2}\,\sigma_{+}\sigma_{-}\,\rho_t -\frac{1}{2}\, \rho_t\,\sigma_{+}\sigma_{-}\right),
\end{split}
\end{equation}

\noindent where $\sigma_{\pm}=\sigma_x\pm i\sigma_y$ are the Pauli  operators and $\gamma_t$ is the time-dependent decay rate given by,

\begin{equation}
\label{eq34}
\gamma_t = \frac{2\gamma_0 \lambda \sinh(h_1t/2)}{h_1\cosh(h_1t/2) + \lambda \sinh(h_1t/2)},
\end{equation}

\noindent where $h_1=\sqrt{\lambda^2-2\gamma_0\lambda}$.


The Bloch representation of a mixed initial quantum state (with a computational basis order $\{\ket{0}, \ket{1}\}$) can be expressed as

\begin{align}
\rho_0=\left(
          \begin{array}{cc}
           1- \rho_{11}(0) & \rho_{01}(0) \\
            \rho_{01}^*(0) & \rho_{11}(0) \\
          \end{array}
        \right)=\frac{1}{2}\left(
                  \begin{array}{cc}
                    1+r_z & r_x-ir_y \\
                    r_x+ir_y & 1-r_z \\
                  \end{array}
                \right),\label{eq:Bloch-rho0}
\end{align}

\noindent where, $r_x,r_y,r_z$ are the components of the Bloch vector.\par
The state of the quantum system at time $t$ in this model is analytically given by~\cite{breuer07, Breuer1999}

\begin{align}
\rho_t =\left(\begin{array}{cc}
1-\rho_{11}(0)\vert q_t\vert^{2} & \rho_{01}(0)q_t\\
\rho_{10}(0)q_t^{*} & \rho_{11}(0)\vert q_t\vert^{2}
\end{array}\right)
\label{eq:dJC-rhot}
\end{align}

\noindent where

\begin{equation}
q_t=
e^{-\lambda t/2}\!\times\!
\begin{cases}
\displaystyle \cosh\!\Big(\frac{h_1 t}{2}\Big)+\frac{\lambda}{h_1}\sinh\!\Big(\frac{h_1 t}{2}\Big),
& (\gamma_0\le \lambda/2),\\[1.0ex]
\displaystyle \cos\!\Big(\frac{h_2 t}{2}\Big)+\frac{\lambda}{h_2}\sin\!\Big(\frac{h_2 t}{2}\Big),
& (\gamma_0> \lambda/2),
\end{cases}
\label{eq:q_t}
\end{equation}

\noindent with $h_2 = \sqrt{2\gamma_0\lambda-\lambda^{2}}$.\par

The time derivative of the density matrix $\rho_t$ can be written as:

\begin{equation}
\dot{\rho_t} = \begin{pmatrix}
-2\rho_{11}(0)  q_t \dot{q}_t & \rho_{01}(0)\dot{q}_t \\
\rho_{10}(0)\dot{q}_t & 2\rho_{11}(0) q_t \dot{q}_t
\end{pmatrix}
\end{equation}

Because $\dot\rho_t$ is Hermitian (off–diagonal entries are conjugates and diagonal entries are real), its \textit{singular values} are the \textit{absolute values of its eigenvalues}. In this case, both singular values turn out to be equal and given by:

\begin{align*}
s_1=s_2= &|\dot q_t|\;\sqrt{\,\big(2\rho_{11}(0)q_t\big)^2+\big|\rho_{01}(0)\big|^2\,}.
\end{align*}

The operator (spectral) norm is the \textit{largest singular value}, thus, in terms of the components of the Bloch vector of the initial state $\rho_0$ we can write:

\begin{align}
\|\dot\rho_t\|_{\mathrm{op}}
=&\frac{|\dot q_t|}{2}\,\sqrt{\,r_x^2+r_y^2+4(1-r_z)^2 q_t^{\,2}\,}.
\end{align}

As the trace norm is the sum of singular values we have $\|\dot\rho_t\|_1 = 2s_1$. Furthermore, Hilbert–Schmidt norm $\|\dot\rho_t\|_2=\sqrt{\Tr{\dot\rho_t^\dagger\dot\rho_t}}=\sqrt{s_1^2+s_2^2}=\sqrt{2}\,s_1$. Thus, 
these three norms are all proportional to the same geometric factor
$\sqrt{\,r_x^2+r_y^2+4(1-r_z)^2 q_t^{2}\,}$, and we have: $\|\dot\rho_t\|_1 = 2\|\dot\rho_t\|_{\mathrm{op}}$, and $\|\dot\rho_t\|_2=\sqrt{2}\|\dot\rho_t\|_{\mathrm{op}}$

\subsection{Bures fidelity}
\label{ssec:buresfidelity}
Bearing in mind Ref.~\cite{Deffner2013} we shall examine the case where the system is initially in the excited state, i.e., $\rho_0 = \ketbra{1}{1}$.
In this case, from equations \eqref{eq:genMLDeffLindbladevol}, \eqref{eq:MTboundDeffLindbladevol}, \eqref{eq:Bloch-rho0}, and \eqref{eq:dJC-rhot} we can write

\begin{align}
\tau_{\text{\tiny QSL}}^{B}
=\frac{1-|q_\tau|^2}{\frac{2\mu}{\tau}\int_0^\tau \,dt\,|q_t||\dot{q}_t|},
\label{eq:BuresQSL}
\end{align}
%

\noindent where $\mu=1,\sqrt{2},2$ for the operator, Hilbert-Schmidt and trace norm, respectively.\par

\begin{figure}[h!]
\includegraphics[width=0.45\textwidth]{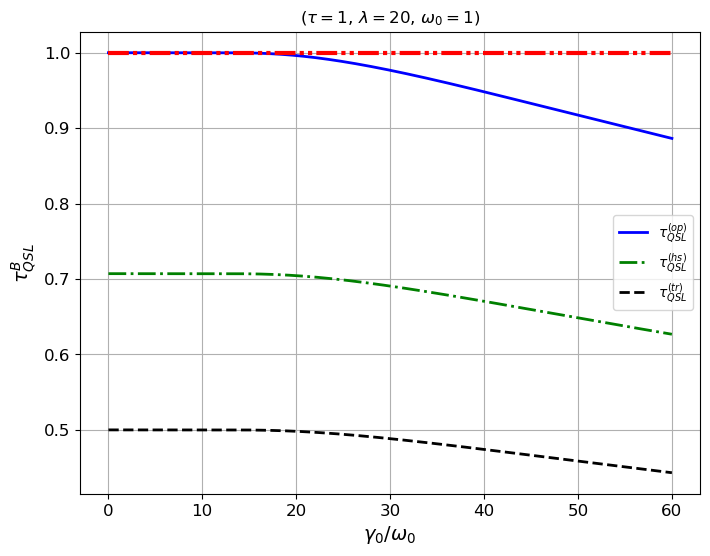}
\caption{ML and MT QSLs $\tau_{\text{\tiny QSL}}^B$ given by Eq. \eqref{eq:BuresQSL}, as a function of $\gamma_0/\omega_0$ for the excited initial state given by $\rho_0 = \ketbra{1}{1}$. The evolution time was set as $\tau=1$, $\lambda = 20$, and $\omega_0=1$. The horizontal dashed line represents the actual driving time $\tau=1$.}
\label{fig:FBresults2Dlamb20}
\end{figure}

\begin{figure}[h!]
\centering
\includegraphics[width=0.45\textwidth]{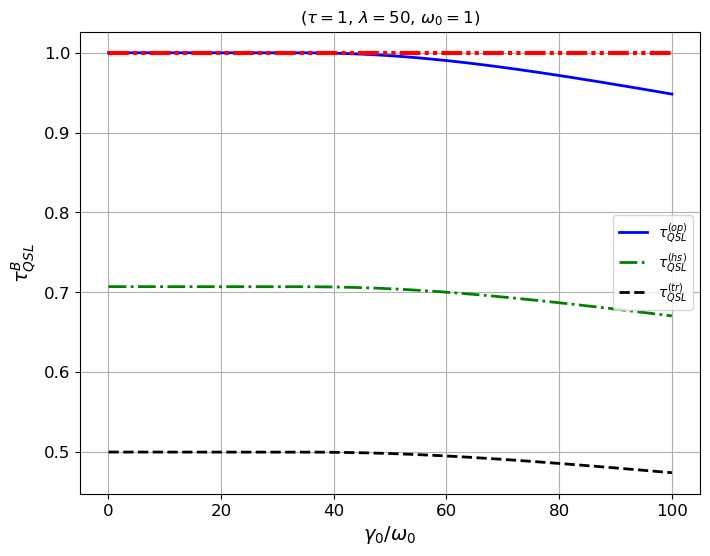}
\caption{ML and MT QSLs $\tau_{\text{\tiny QSL}}^B$ given by Eq. \eqref{eq:BuresQSL}, as a function of $\gamma_0/\omega_0$ for the excited initial state given by $\rho_0 = \ketbra{1}{1}$. The evolution time was set as $\tau=1$, $\lambda = 50$, and $\omega_0=1$. The horizontal dashed line represents the actual driving time $\tau=1$.}
\label{fig:FBresults2Dlamb50}
\end{figure}

Figures \ref{fig:FBresults2Dlamb20} and \ref{fig:FBresults2Dlamb50} show the quantum speed limit $\tau_{\text{\tiny QSL}}^B$, given by equation \eqref{eq:BuresQSL}, evaluated for the three different norms as a function of $\gamma_0/\omega_0$, for $\tau=1$, and for $\lambda=20$ and $\lambda=50$, respectively. In agreement with equations \eqref{eq:genMLDeffLindbladevol} and \eqref{eq:MTboundDeffLindbladevol}, the MT-type QSL obtained from the Hilbert–Schmidt and the ML-type QSL obtained with the trace norm are looser than the ML-type QSL derived from the operator norm.
As shown in the figures, $\tau_{\text{\tiny QSL}}^B$ exhibits a plateau over a relatively wide range of the parameter $\gamma_0/\omega_0$, after which the QSL decreases for larger values of $\gamma_0/\omega_0$.
Moreover, the ML QSL corresponding to the operator norm becomes tight over a broad interval of $\gamma_0/\omega_0$, since it attains the actual evolution time $\tau$ within that range.


\subsection{Super-fidelity}
\label{ssec:Super-fidelity}

Assuming that the system is in the initial state $\rho_0$ given by equation \eqref{eq:Bloch-rho0}, and using Eq. \eqref{eq:MLboundsuperfidelity}, we can write the  ML-type QSL as

\begin{align*}
\tau_{\text{\tiny QSL}}^{s}
=\frac{\tfrac12\Big[1-r_z-q_\tau(r_x^2+r_y^2)+|q_\tau|^{2}r_z(1-r_z)-\kappa_1\,\kappa_2^{\,\tau}\Big]}
{\dfrac{1}{\tau}\!\int_0^\tau \! dt\;\dfrac{|\dot q_t|}{2}\,
\sqrt{\,r_x^2+r_y^2+4(1-r_z)^2 |q_t|^{2}\,}\;
\Big(1+\dfrac{\kappa_1}{\kappa_2^{\,t}}\Big)}.
\end{align*}

\noindent where

\begin{align*}
\kappa_1&\doteq\sqrt{1-r_x^2-r_y^2-r_z^2},\\ 
\kappa_2^{\,t}&\doteq\sqrt{\,|q_t|^2\!\Big(2(1-r_z)-(r_x^2+r_y^2) - |q_t|^2(1-r_z)^2\Big)}.
\end{align*}

If the initial state is taken as the excited state $\ketbra{1}$ with added white noise, i.e., $\rho_0=\tfrac{p}{2}\mathbb I+(1-p)\,|1\rangle\!\langle 1|$,
then in the basis $\{\ket{0},\ket{1}\}$ $\tau_{\text{\tiny QSL}}^s$ can be written as

\begin{equation}
\tau_{\text{\tiny QSL}}^{s}=\frac{\tfrac12\left[(2-p)\Big(1-(1-p)q_\tau^{2}\Big) - \tilde{\kappa}_1\,\tilde{\kappa}_2^{\,\tau}\right]}
{\frac{1}{\tau}\!\int_0^\tau \! dt\;\;(2-p)\,|q_t|\,|\dot q_t|
\Big(1+\frac{\tilde{\kappa}_1}{\tilde{\kappa}_2^{\,t}}\Big)}
\label{eq:superfidelMLQSL}
\end{equation}

where

\begin{align*}
\tilde{\kappa}_1=&\sqrt{p\,(2-p)},\\
\tilde{\kappa}_2^{\,t}=&\sqrt{(2-p)\,|q_t|^{2}\left[\,2-(2-p)\,|q_t|^{2}\right]}.
\end{align*}

\begin{figure}[h!]
    \includegraphics[width=0.5\textwidth]{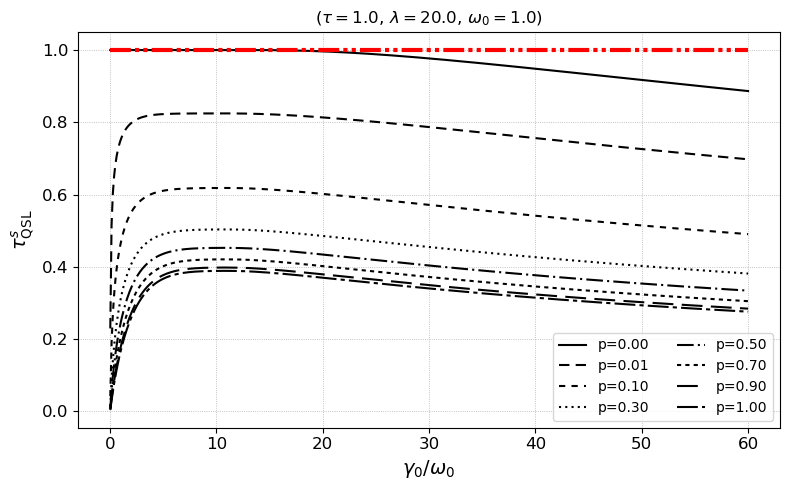}
    \caption{ML-type QSL $\tau_{\text{\tiny QSL}}^s$ given by Eq. \eqref{eq:superfidelMLQSL}, as a function of $\gamma_0/\omega_0$ for different initial states of the form $\rho_0=(p/2)\mathbb I+(1-p)\,|1\rangle\!\langle 1|$. The evolution time was set as $\tau=1$, $\lambda = 20$, and $\omega_0=1$. The horizontal dashed line represents the actual driving time $\tau=1$.}
    \label{fig:Fsresults2Dlamb20}
\end{figure}

\begin{figure}[h!]
    \includegraphics[width=0.5\textwidth]{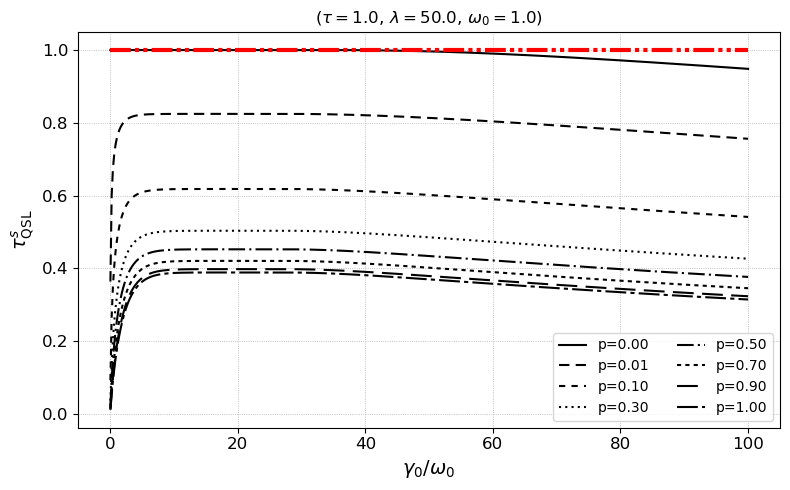}
    \caption{ML-type QSL $\tau_{\text{\tiny QSL}}^s$ given by Eq. \eqref{eq:superfidelMLQSL}, as a function of  $\gamma_0/\omega_0$ for different initial states of the form $\rho_0=(p/2)\mathbb I+(1-p)\,|1\rangle\!\langle 1|$. The evolution time was set as $\tau=1$, $\lambda = 50$, and $\omega_0=1$. The horizontal dashed line represents the actual driving time $\tau=1$.}
    \label{fig:Fsresults2Dlamb50}
\end{figure}

Figures \ref{fig:Fsresults2Dlamb20} and \ref{fig:Fsresults2Dlamb50} show the ML-type QSL $\tau_{\text{\tiny QSL}}^s$, given by equation \eqref{eq:superfidelMLQSL}, evaluated for the operator norm as a function of $\gamma_0$, for $\tau=1$, and for $\lambda=20$ and $\lambda=50$, respectively ($\gamma_0$ is set in units of $\omega_0$). In both figures, it can be observed that the ML-type QSL $\tau_{\text{\tiny QSL}}^s$ is not tight for $0<p\leq 1$, as it never reaches the actual driving time $\tau=1$.
For very small values of $\gamma_0$ and $0<p\leq1$, a rapid increase in $\tau_{\text{\tiny QSL}}^s$ with a steep slope is observed at the beginning, until a plateau is reached for moderate values of $\gamma_0$. For larger values of the coupling constant, $\tau_{\text{\tiny QSL}}^s$ decreases. As expected, since $\mathcal{F}_s(\rho_0,\rho_\tau)$ reduces to the Bures fidelity $F_B(\rho_0,\rho_\tau)$ for pure initial states, when $p=0$ (i.e., $\rho_0 = \ketbra{1}{1}$) the ML-type QSL $\tau_{\text{\tiny QSL}}^s$ coincides with the ML-type QSL $\tau_{\text{\tiny QSL}}^B$ corresponding to the operator norm, as shown in Figs.~\ref{fig:FBresults2Dlamb20} and \ref{fig:FBresults2Dlamb50}.

\subsection{Operator fidelity}
\label{ssec:operatorfidelity}

Assuming that the system is in the initial state $\rho_0$ given by equation \eqref{eq:Bloch-rho0}, and using Eq. \eqref{eq:MTQSLoperfidel}, we can write the  MT-type QSL as

\begin{equation*}
\tau_{\text{\tiny QSL}}^o
=\frac{
\displaystyle 1-\frac{1+\vec r\cdot \vec r(\tau)}
{\sqrt{(1+|\vec r|^2)(1+|\vec r(\tau)|^2)}}
}{
\displaystyle \frac{2}{\tau}\int_0^\tau
\frac{|\dot q_t|
\sqrt{r_x^2+r_y^2+4(1-r_z)^2\,|q_t|^{2}\,}}
{\sqrt{1+|q_t|^{2}(r_x^2+r_y^2)+\big[1-(1-r_z)|q_t|^{2}\big]^2\,}}
\;dt }.
\label{eq:qubitMToperfidel}
\end{equation*}

where

\begin{align*}
|\vec r|^2=&r_x^2+r_y^2+r_z^2,\\
\vec r\cdot \vec r(\tau)=&
q_\tau(r_x^2+r_y^2)+r_z-(r_z-r_z^2)|q_\tau|^2,\\
|\vec r(\tau)|^2=& |q_\tau|^2(r_x^2+r_y^2)+\big[1-(1-r_z)|q_\tau|^2\big]^2.
\end{align*}

If the initial state is taken as $\rho_0=\tfrac{p}{2}\mathbb I+(1-p)\,|1\rangle\!\langle 1|$, then in this case $\tau_{\text{\tiny QSL}}^o$ can be written as

\begin{equation}
\tau_{\text{\tiny QSL}}^o=
\frac{\displaystyle
1-
\frac{1+(p-1)\big[1-(2-p)q_\tau^2\big]}
{\sqrt{\big(1+(1-p)^2\big)\big(1+\big[1-(2-p)q_\tau^2\big]^2\big)}}
}
{\displaystyle
\frac{2}{\tau}\int_0^\tau
\frac{2(2-p)|q_t\dot q_t|}
{\sqrt{1+\big[1-(2-p)q_t^2\big]^2}}\;dt
}
\;.
\label{eq:operfidelMTQSL}
\end{equation}

\begin{figure}[h!]
    \includegraphics[width=0.5\textwidth]{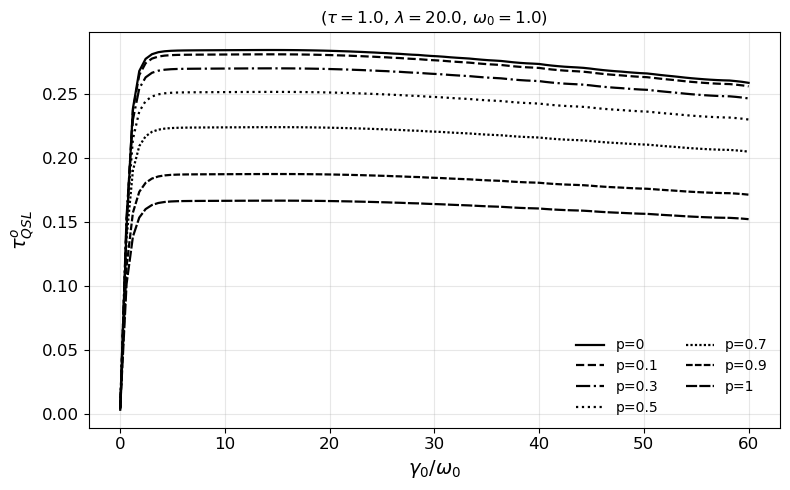}
    \caption{MT-type QSL $\tau_{\text{\tiny QSL}}^o$ given by Eq.\eqref{eq:operfidelMTQSL}, as a function of $\gamma_0/\omega_0$ for different  initial states of the form $\rho_0=(p/2)\mathbb I+(1-p)\,|1\rangle\!\langle 1|$. The evolution time was set as $\tau=1$, $\lambda = 20$, and $\omega_0=1$.}
    \label{fig:Foresults2Dlamb20}
\end{figure}

\begin{figure}[h!]
    \includegraphics[width=0.5\textwidth]{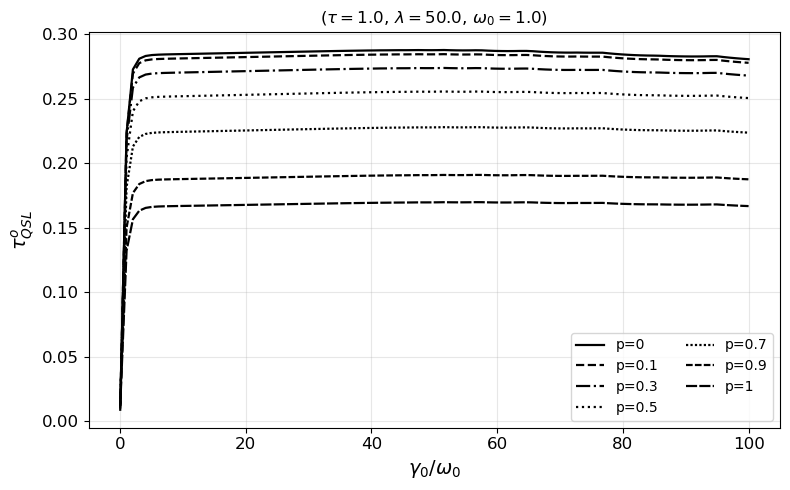}
    \caption{MT-type QSL $\tau_{\text{\tiny QSL}}^o$ given by Eq.\eqref{eq:operfidelMTQSL}, as a function of $\gamma_0/\omega_0$ for different initial states of the form $\rho_0=(p/2)\mathbb I+(1-p)\,|1\rangle\!\langle 1|$. The evolution time was set as $\tau=1$, $\lambda = 50$, and $\omega_0=1$.}
    \label{fig:Foresults2Dlamb50}
\end{figure}

Figures \ref{fig:Foresults2Dlamb20} and \ref{fig:Foresults2Dlamb50} show the MT-type QSL $\tau_{\text{\tiny QSL}}^o$, given by equation \eqref{eq:operfidelMTQSL}, as a function of $\gamma_0$, for $\tau=1$, and for $\lambda=20$ and $\lambda=50$, respectively. In both figures, it can be observed that the MT-type QSL $\tau_{\text{\tiny QSL}}^o$ is not tight as it never reaches the actual driving time $\tau=1$. As in the case of the ML-type QSL $\tau_{\text{\tiny QSL}}^s$, for very small values of $\gamma_0$ and $0<p\leq1$, a rapid increase in $\tau_{\text{\tiny QSL}}^o$ with a steep slope is observed initially, until a plateau is reached for moderate values of $\gamma_0$. For larger values of the coupling constant $\gamma_0$, $\tau_{\text{\tiny QSL}}^o$ decreases. In this case, it is worth noting that $\mathcal{F}_o(\rho_0,\rho_\tau)$ does not reduce to the Bures fidelity $F_B(\rho_0,\rho_\tau)$ for pure initial states, due to the additional factor $1/\sqrt{\Tr{\rho_t^2}}$ (cf.~Eq.~\eqref{eq:operfidel}). Therefore, for $p=0$ (i.e., $\rho_0=\ketbra{1}{1}$), $\tau_{\text{\tiny QSL}}^o$ does not coincide with the $\tau_{\text{\tiny QSL}}^B$ curves corresponding to the Hilbert–Schmidt norm, shown in Figs.~\ref{fig:FBresults2Dlamb20} and \ref{fig:FBresults2Dlamb50}.

\subsection{Alternative fidelity}
\label{ssec:alterfidelity}

We shall assume the system to be in the initial state $\rho_0$ given by equation \eqref{eq:Bloch-rho0}. Using Eq. \eqref{eq:MTQSLektesabifidel}, the MT-type QSL can be written as

\[
\tau_{\text{\tiny QSL}}^a=\frac{1-\frac12\bigl(1+g_0 g_\tau\bigr)\bigl[1+\bm r\cdot\bm R_\tau\bigr]}{\frac{1}{\tau}\int_0^\tau \bigl[T_1(t)+T_2(t)+T_3(t)\bigr]\,dt},
\]

\noindent where

\begin{align*}
  g_0 & \doteq \sqrt{\frac{1-r^2}{1+r^2}},\\[2pt]
  g_t & \doteq \sqrt{\frac{1-R(t)^2}{1+R(t)^2}},\\[2pt]
  \bm r & \doteq (r_x,r_y,r_z),\\[2pt]
  \bm R(t)
  & \doteq\bigl(r_x q_t,\; r_y q_t,\; 1-(1-r_z)q_t^2\bigr),\\[2pt]
 T_1(t)
  &\doteq g_0\,g_t^{-1}\,
    \left|
    \frac{\bigl[\dot{\bm R}(t)\cdot\bm R(t)\bigr]
          \bigl[1+\bm r\cdot\bm R(t)\bigr]}
         {\bigl[1+R(t)^2\bigr]^2}
    \right|,\\[2pt]
T_2(t)
  &\doteq \sqrt{\frac14(1+r^2)\,|\dot{\bm R}(t)|^2},\\[2pt]
  T_3(t)
  &\doteq g_t\sqrt{\frac14(1-r^2)\,|\dot{\bm R}(t)|^2}.
  \end{align*}
 
\noindent and we define $q_\tau \doteq q_t\big|_{t=\tau}$, $\bm R_\tau \doteq \bm R(\tau)$, and $g_\tau \doteq g_t\big|_{t=\tau}$.\par

 As before, we shall assume the initial state to be $\rho_0=\tfrac{p}{2}\mathbb I+(1-p)\,|1\rangle\!\langle 1|$. In this case $\tau_{\text{\tiny QSL}}^a$ can be written as

\begin{equation}
\tau_{\text{\tiny QSL}}^a = \frac{1 - \frac12\bigl(1+\tilde{g}_0 \tilde{g}_\tau\bigr)\bigl[1+r_z R_\tau\bigr]}{\frac{1}{2\tau}\int_0^\tau |\dot R_z(t)|\,K(R_z(t);p)\,dt},
\label{eq:alterfidelMTQSL}
\end{equation}

\noindent with $r_z=p-1$, $R_\tau=1-(2-p)q_\tau^2$,

\begin{equation*}
  \tilde{g}_0 = \sqrt{\frac{1-r_z^2}{1+r_z^2}},\qquad
  \tilde{g}_t = \sqrt{\frac{1-R_z(t)^2}{1+R_z(t)^2}},
\end{equation*}

\noindent and the weight function $K(R_z;p)$ is given by

\begin{align}
  K(R_z;p)
  &= \tilde{g}_0\,g(R_z)^{-1}\,
     \frac{2\,\bigl|R_z\,[1+r_z R_z]\bigr|}
          {\bigl[1+R_z^2\bigr]^2}
   +\nonumber\\[2pt]
   &\sqrt{1+r_z^2}
   + g(R_z)\,\sqrt{1-r_z^2},
  \label{eq:K_def}
\end{align}

\noindent with

\begin{equation*}
 g(R_z) = \sqrt{\frac{1-R_z^2}{1+R_z^2}}.
\end{equation*}


\begin{figure}[h!]
    \includegraphics[width=0.5\textwidth]{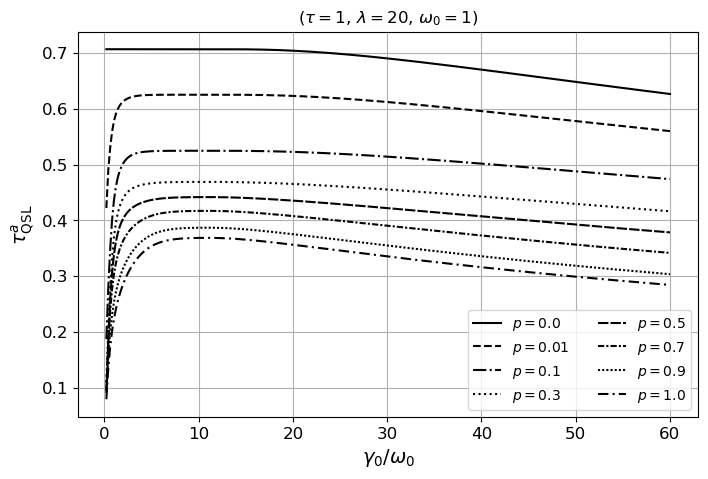}
    \caption{MT-type QSL $\tau_{\text{\tiny QSL}}^a$ given by Eq.\eqref{eq:alterfidelMTQSL}, as a function of $\gamma_0/\omega_0$ for different initial states of the form $\rho_0=(p/2)\mathbb I+(1-p)\,|1\rangle\!\langle 1|$. The evolution time was set as $\tau=1$, $\lambda = 20$ and $\omega_0 = 1$.}
    \label{fig:alterfidel2Dplotlamb20}
\end{figure}

\begin{figure}
    \includegraphics[width=0.5\textwidth]{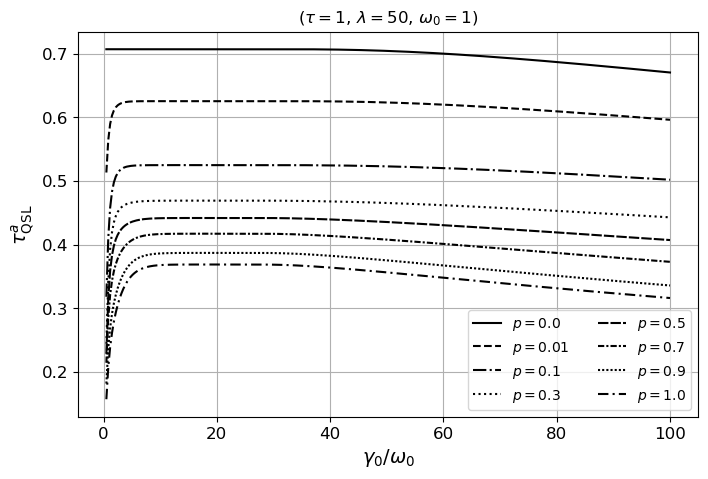}
    \caption{MT-type QSL $\tau_{\text{\tiny QSL}}^a$ given by Eq.\eqref{eq:alterfidelMTQSL}, as a function of $\gamma_0/\omega_0$ for different initial states of the form $\rho_0=(p/2)\mathbb I+(1-p)\,|1\rangle\!\langle 1|$. The evolution time was set as $\tau=1$, $\lambda = 50$ and $\omega_0 = 1$.}
    \label{fig:alterfidel2Dplotlamb50}
\end{figure}

Figures \ref{fig:alterfidel2Dplotlamb20} and \ref{fig:alterfidel2Dplotlamb50} show the MT-type QSL $\tau_{\text{\tiny QSL}}^a$, given by equation \eqref{eq:alterfidelMTQSL}, as a function of $\gamma_0/\omega_0$, for $\tau=1$, and for $\lambda=20$ and $\lambda=50$, respectively. It can be observed that the MT-type QSL $\tau_{\text{\tiny QSL}}^a$ is not tight as it never reaches the actual driving time $\tau=1$. As in previous cases, for very small values of $\gamma_0$ and $0<p\leq1$, a rapid increase in $\tau_{\text{\tiny QSL}}^a$ with a steep slope is observed initially, until a plateau is reached for moderate values of $\gamma_0$. For larger values of the coupling constant $\gamma_0$, $\tau_{\text{\tiny QSL}}^a$ decreases. As expected, since $\mathcal{F}_a(\rho_0,\rho_\tau)$ reduces to the Bures fidelity $F_B(\rho_0,\rho_\tau)$ for pure initial states, $\tau_{\text{\tiny QSL}}^a$ coincides with the $\tau_{\text{\tiny QSL}}^B$ curves corresponding to the Hilbert–Schmidt norm when $p=0$, as shown in Figs.~\ref{fig:FBresults2Dlamb20} and \ref{fig:FBresults2Dlamb50}.

\section{Analysis and discussion}
\label{sec:analysis-discussion}
\subsection{Analytical results}

Despite the apparent freedom in the choice of the function \( D(x) \) mentioned in Proposition 1 and in Theorems 1 through 6, it can be observed that all the derived QSLs ultimately depend only on the chosen fidelity and are independent of the specific form of \( D(x) \), provided that \( D(x) \) satisfies the properties stated in Proposition 1.

In the case of the Bures fidelity $F_B$ Eq.~\eqref{eq:genMLDeffHamiltdyn} provides a ML-type QSL for closed quantum systems under time-dependent Hamiltonian evolution. In addition, Eq.~\eqref{eq:genMLDeffLindbladevol} generalizes the ML-type QSL to arbitrary open system dynamics generated by positive time-local maps.
Combining Eqs.~\eqref{eq:genMLDeffLindbladevol} and \eqref{eq:MTboundDeffLindbladevol}, we obtain the unified QSL for Bures fidelity $F_B$ as follows:

\begin{equation*}
\tau_{\mbox{\tiny QSL}} = \max \left\{ \frac{1}{\mathcal{B}^{\mbox{\tiny (op)}}(\tau)}, \frac{1}{\mathcal{B}^{\mbox{\tiny (Tr)}}(\tau)}, \frac{1}{\mathcal{B}^{\mbox{\tiny (HS)}}(\tau)} \right\} \left[ 1 - F_B(\tau) \right],
\end{equation*}

\noindent which generalizes both the ML-type and MT-type QSLs to arbitrary open-system dynamics.\par

In the case of super fidelity, by combining Eqs.~\eqref{eq:MLboundsuperfidelity} and \eqref{eq:MTboundsuperfidelity} we obtain n unified expression for the QSL as follows:

\begin{align*}
\tau_{\mbox{\tiny QSL}} = \max \left\{ \frac{1}{\mathcal{B}^{\mbox{\tiny (op)}}(\tau)}, \frac{1}{\mathcal{B}^{\mbox{\tiny (Tr)}}(\tau)}, \frac{1}{\mathcal{B}^{\mbox{\tiny (HS)}}(\tau)} \right\} \left[ 1 - \mathcal{F}_s(\tau) \right].
\end{align*}

Furthermore, the following inequality for trace-class operators holds~\cite{sim05}:

\begin{equation}
\label{eq24}
\| A \|_{\mbox{\tiny op}} \leq \| A \|_{\mbox{\tiny HS}} \leq \| A \|_{\mbox{\tiny Tr}}.
\end{equation}

\noindent Therefore, $1 / \mathcal{B}^{\mbox{\tiny (op)}}(\tau) \geq 1 / \mathcal{B}^{\mbox{\tiny (HS)}}(\tau) \geq 1 / \mathcal{B}^{\mbox{\tiny (Tr)}}(\tau)$, pointing out that the ML-type QSL using the operator norm should provide the sharpest estimate for the QSL time in all cases.

The Bures angle between a pure initial state and a final state represented by a general density operator, is given by:

\begin{equation}
\label{eq:buresangle}
\mc{L}(\rho_0,\rho_t)=\arccos\left(\sqrt{\bra{\psi_0}\rho_t\ket{\psi_0}}\right),
\end{equation}

\noindent where

\[
F_{B}(\rho_{0},\rho_{t}) =\bra{\psi_0}\rho_t\ket{\psi_0}
\]

\noindent can be identified as the Bures fidelity between an initial pure state $\rho_0 = \ketbra{\psi_0}$ and a general quantum state represented by the density operator $\rho_t$.

In a seminal work, Deffner et al.~\cite{Deffner2013} used a geometric approach based on the Bures angle (see  Eq.~\eqref{eq:buresangle}) to derive MT and ML-type QSLs. They started their calculations by taking the time derivative of Eq.~\eqref{eq:buresangle}. Then, using von Neumann ~\cite{vonNeumann1937,Mirsky1975,Grigorieff1991,Deffner2013} and Cauchy-Schwarz~\cite{delCampo2013,Deffner2013} inequalities for operators, they arrived at a set of MT and ML-type QSLs. All of the obtained QSLs are written in terms of the Bures angle~\cite{Deffner2013}. However, it is worth mentioning that the metric character of the Bures angle is not needed at any point in their QSL derivation. Furthermore, working out all the QSLs obtained in Ref.~\cite{Deffner2013}, it can be seen that they can be rewritten in terms of Bures fidelity alone. On the other hand, all of their results can be reinterpreted and rewritten to express them in the form given by Theorem 2. Thus, all the QSLs obtained in Ref.~\cite{Deffner2013} are contained within the framework introduced in our work. It is important to note again that the QSLs obtained in Ref.~\cite{Deffner2013} are independent of the Bures angle and any other function of Bures fidelity, $D(F_B)$, that satisfies the properties stated in Proposition~\ref{proposition} and Theorem~\ref{theo:buresfidelity}.\par

An approach similar to that used in Ref.~\cite{Deffner2013} is used by Wu et al. in Ref.~\cite{Wu2020} where the so-called \textit{modified Bures angle} is introduced:\par

\begin{eqnarray}
\mathcal{L}_s(\rho_{0},\rho_t)=\arccos
\left(\sqrt{\mathcal{F}_s(\rho_0,\rho_t)}\right).\label{eq:modifiedburesangle}
\end{eqnarray}

\noindent In last equation, $\mathcal{F}_s$ represents the \textit{super fidelity} given by Eq.~\eqref{eq:superfidelity}.\par

By taking the time derivative of Eq.~\eqref{eq:modifiedburesangle} and using von Neumann and Cauchy inequalities for operators~\cite{vonNeumann1937,Mirsky1975,Grigorieff1991,Deffner2013,delCampo2013,Deffner2013}, Wu et al.~\cite{Wu2020} arrived at a set of MT and ML-type QSLs.

All of the obtained QSLs are written in terms of the quantity $\mathcal{L}_s$~\cite{Wu2020}. Again,  all the QSLs obtained in Ref.~\cite{Wu2020} can be worked out, reinterpreted and rewritten to express them in the form given by Theorem~\ref{theo:superfidelity}. Thus, all the QSLs obtained in Ref.~\cite{Wu2020} are contained within the framework introduced in our work and, additionally, they are independent of the particular choice of $\mathcal{L}_s$  and any other function $D(\mathcal{F}_s)$ of the relative purity (cf.~Eq.~\eqref{eq:superfidelity}) that satisfies the properties stated in Proposition~\ref{proposition} and Theorem~\ref{theo:superfidelity}.\par

\subsection{Numerical results}

\subsubsection{Bures fidelity}

In order to analyze the behavior of $\tau_{\text{\tiny QSL}}^B$ observed in Figs.~1 and 2, we shall examine equation \eqref{eq:BuresQSL} for different values of the coupling constant $\gamma_0$. 

Let us define the excited-state population \(P_t\doteq|q_t|^2\).
%
Then, the denominator of~\eqref{eq:BuresQSL} can be written as a \emph{total variation} (i.e.\ path length) of \(P_t\) on \([0,\tau]\):

\begin{equation*}
\frac{2\mu}{\tau}\int_0^\tau |q_t|\,|\dot q_t|\,dt
= \frac{\mu}{\tau}\int_0^\tau |\dot P_t|\,dt.
\end{equation*}

The numerator is simply the net decay of the excited population over \([0,\tau]\): $1-|q_\tau|^2 = 1-P_\tau.$ Hence

\begin{equation}
\tau_{\text{\tiny QSL}}^{B}
= \frac{\tau}{\mu}\;
\frac{\,1-P_\tau\,}{\displaystyle\int_0^\tau |\dot P_t|\,dt}.
\label{eq:QSL_master}
\end{equation}

Equation~\eqref{eq:QSL_master} shows that the qualitative behavior of \(\tau_{\text{\tiny QSL}}^{B}\) is entirely controlled by the relation between the \emph{net drop} \(1-P_\tau\) and the \emph{accumulated path length} \(\int_0^\tau |\dot P_t|\,dt\).

\subsubsection*{Markovian regime: weak coupling ($0<\gamma_0\leq \lambda/2$)}

When \(\gamma_0\le \lambda/2\), Eq.~\eqref{eq:q_t} implies \(q_t>0\) and
\(
\dot q_t\le 0
\)
for \(t>0\).
Hence \(P_t=|q_t|^2\) is \emph{strictly decreasing} on \([0,\tau]\), and therefore

\begin{equation}
\int_0^\tau |\dot P_t|\,dt
= \int_0^\tau \big(-\dot P_t\big)\,dt
= P_0-P_\tau
= 1-P_\tau.
\label{eq:TV_equals_drop}
\end{equation}

Inserting~\eqref{eq:TV_equals_drop} into~\eqref{eq:QSL_master} we obtain the \emph{exact plateau}:

\begin{equation}
\tau_{\text{\tiny QSL}}^{B} = \frac{\tau}{\mu}.
\label{eq:plateau_all_norms}
\end{equation}

Thus, for fixed values of \(\tau\) and \(\lambda\), the quantity \(\tau_{\text{\tiny QSL}}^{B}\) remains constant throughout the Markovian region, taking the values \(\tau_{\text{\tiny QSL}}^{B} = \tau\), \(\tau_{\text{\tiny QSL}}^{B} = \tau/\sqrt{2}\), and \(\tau_{\text{\tiny QSL}}^{B} = \tau/2\), for the operator norm, Hilbert--Schmidt norm, and trace norm, respectively. Physically, the state follows a \emph{monotone} path from \(P_0=1\) to \(P_\tau\); the geometric ``distance traveled'' (denominator) equals the net drop (numerator) up to the prefactor \(\mu/\tau\), so their ratio is the constant~\eqref{eq:plateau_all_norms}.
This explains the observed \emph{plateau} for \(\gamma_0<\lambda/2\).

\subsubsection*{non-Markovian regime: strong coupling ($\gamma_0>\lambda/2$)}

For \(\gamma_0>\lambda/2\), the amplitude is oscillatory under a decaying envelope:

\begin{align*}
q_t
=& e^{-\lambda t/2}\!\left[\cos\!\Big(\frac{h_2 t}{2}\Big)+\frac{\lambda}{h_2}\sin\!\Big(\frac{h_2 t}{2}\Big)\right]\\
=& A\,e^{-\lambda t/2}\cos\!\Big(\frac{h_2 t}{2}-\delta\Big),
\label{eq:osc_form}
\end{align*}

\noindent with \(A=\sqrt{1+(\lambda/h_2)^2}\) and \(\delta=\arctan(\lambda/h_2)\in(0,\pi/2)\).
Consequently, \(P_t=|q_t|^2\) is \emph{not} monotonic. The effective decay rate becomes time-dependent and oscillatory. Physically, information may transiently flow back from the environment to the system.


Because \(P_t\) is not monotonic, its \emph{total variation} strictly exceeds the net drop:

\begin{equation*}
\int_0^\tau |\dot P_t|\,dt \;>\; P_0-P_\tau \;=\; 1-P_\tau,
\end{equation*}

\noindent whenever \(P_t\) has at least one local increase on \([0,\tau]\). From~\eqref{eq:QSL_master} this implies the strict inequality

\begin{equation*}
\tau_{\text{\tiny QSL}}^{B} < \frac{\tau}{\mu},
\end{equation*}

\noindent for $\gamma_0>\lambda/2$, and at least one revival of $P_t \text{ in } (0,\tau)$. This means that \emph{non-Markovian oscillations} increase the path length in state space (the integral of \(|\dot P_t|\)), while the net drop \(1-P_\tau\) remains bounded by the envelope. Thus, their ratio \emph{decreases} below the plateau.

\subsubsection{Super fidelity}

In order to analyze the behavior of $\tau_{\text{\tiny QSL}}^s$ observed in Figs.~3 and 4, we shall examine equation \eqref{eq:BuresQSL} for different values of the coupling constant $\gamma_0$.\par

For the ML--type QSL with super-fidelity \(\mathcal F_s\), we can write

\begin{align*}
\tau_{\text{\tiny QSL}}^s=&\frac{\text{N}_s}{\text{D}_s},
\end{align*}

\noindent where

\begin{align*}
\text{N}_s=&1-\mathcal F_s(\rho_0,\rho_\tau),\\
\text{D}_s=&\frac{1}{\tau}\!\int_0^\tau\!\|\dot\rho_t\|_{\rm op}\!\left(1+\sqrt{\frac{1-\Tr{\rho_0^2}}{1-\Tr{\rho_t^2}}}\right)\!dt.
\end{align*}

In this setting one obtains the compact forms

\begin{align*}
\text{N}_s
&=\frac12\Big[(2-p)\big(1-(1-p)q_\tau^2\big)-\sqrt{p(2-p)}\times\\
&\,\sqrt{(2-p)\,q_\tau^2\big(2-(2-p)q_\tau^2\big)}\Big],\label{eq:Num-min}\\
\text{D}_s
&=\frac{1}{\tau}\int_0^\tau (2-p)\,|q_t||\dot q_t|\!\left(1+\sqrt{\frac{p}{q_t^2\big(2-(2-p)q_t^2\big)}}\right)\!dt.
\end{align*}

\subsubsection*{Markovian regime: very weak coupling ($0<\gamma_0\ll\lambda$)}

From the first branch of \eqref{eq:q_t}, a uniform small-\(\gamma_0\) expansion yields

\begin{align*}
q(t)=&1-\frac{\gamma_0}{2}\!\left[t-\frac{1-e^{-\lambda t}}{\lambda}\right]+O(\gamma_0^2),\\
\dot q(t)=&-\frac{\gamma_0}{2}\big(1-e^{-\lambda t}\big)+O(\gamma_0^2).
\end{align*}

Thus \(|q_t|=1+O(\gamma_0)\) and \(|\dot q_t|=O(\gamma_0)\), monotonically increasing on \([0,\tau]\).
Writing \(P_\tau\doteq|q_\tau|^2=1-\varepsilon+O(\gamma_0^2)\) with \(\varepsilon=O(\gamma_0)\), and $p>0$ the super-fidelity numerator satisfies

\begin{equation*}
\text{N}_s= \frac{1}{4}\,(2-p)\!\left[(2-p)+\frac{(1-p)^2}{p}\right]\varepsilon^2+O(\varepsilon^3)
=\;O(\gamma_0^2),
\end{equation*}

\noindent whereas the denominator obeys

\begin{equation*}
\text{D}_s=\frac{(2-p)}{\tau}\int_0^\tau \gamma_0\big(1-e^{-\lambda t}\big)\,dt +O(\gamma_0^2)
=\;O(\gamma_0).
\end{equation*}

Therefore, for any fixed \(p\in(0,1]\),

\begin{equation*}
\tau_{\text{\tiny QSL}}^{s}(\gamma_0)\;=\;c(p,\lambda,\tau)\,\gamma_0+O(\gamma_0^2),
\end{equation*}

\noindent i.e., it \emph{starts at zero and rises linearly} with \(\gamma_0\).
For the special case \(p=0\), the first-order cancellation disappears, \(\text{N}_s=O(\gamma_0)\) and \(\text{D}_s=O(\gamma_0)\),
and in the monotone sector one has the exact identity \(\text{D}_s=\text{N}_s/\tau\), giving

\begin{equation*}
\tau_{\text{\tiny QSL}}^{s}=\tau\quad (p=0,\ \gamma_0\le \lambda/2).
\end{equation*}

\subsubsection*{Markovian regime: moderate coupling ($0<\gamma_0\le\lambda/2$)}

When \(q_t\) decays monotonically and \(q_\tau\!\to\!0\),

\begin{align*}
\text{N}_s\to & \frac{2-p}{2},\\
\text{D}_s\to &\frac{2-p}{\tau}\!\left(\frac12+\frac{\sqrt{p}}{\sqrt{2}}\right),
\end{align*}

so the ratio settles on the plateau

\begin{equation*}
\tau_{\text{\tiny QSL}}^{s}\ \approx\ \frac{\tau}{\,1+\sqrt{2p}\,}\, .
\end{equation*}

\subsubsection*{non-Markovian regime: strong coupling ($\gamma_0>\lambda/2$)}

In this regime \(q_t\) becomes oscillatory and damped; using the second branch of \eqref{eq:q_t} and

\begin{equation*}
\dot q(t)=-\,e^{-\lambda t/2}\,\frac{\gamma_0\lambda}{h_2}\,\sin\!\Big(\tfrac{h_2 t}{2}\Big),
\end{equation*}

the numerator remains near its endpoint-controlled plateau, \(\text{N}_s\approx (2-p)/2\).
In contrast, the denominator contains the \emph{total variation} \(\int_0^\tau |\dot q_t|\,dt\); averaging the fast oscillations with slow envelope gives

\begin{equation*}
\text{D}_s\ \propto\ \frac{\gamma_0}{h_2}\ \sim\ \sqrt{\frac{\gamma_0}{\lambda}}\qquad (\gamma_0\gg\lambda),
\end{equation*}

\noindent yielding a characteristic decline

\begin{equation*}
\tau_{\text{\tiny QSL}}^{s}(\gamma_0)\ \sim\ C(p,\lambda,\tau)\,\sqrt{\frac{\lambda}{\gamma_0}}\ \xrightarrow[\gamma_0\to\infty]{} 0\,.
\end{equation*}

\subsubsection{Operator fidelity}

From Eq. \eqref{eq:operfidelMTQSL}, the operator–fidelity QSL (MT-type) can be written as

\begin{align*}
\tau_{\text{\tiny QSL}}^{o}=&\frac{\text{N}_o}{\text{D}_o},
\end{align*}

\noindent where

\begin{align*}
\text{N}_o\doteq&1-\frac{1+(p-1)u_\tau}{\sqrt{A(1+u_\tau^2)}},\\
\text{D}_o\doteq&\frac{2}{\tau}\int_0^\tau \frac{2(2-p)\,|q_t\dot q_t|}{\sqrt{1+u_t^2}}\,dt.
\end{align*}

We also set

\begin{align*}
u_t\doteq&1-(2-p)\,q_t^2,\\
A\doteq&1+(1-p)^2.
\end{align*}

\subsubsection*{Markovian regime: very weak coupling ($\gamma_0\to0^+$)}

Expanding at fixed time (no short–time assumption),

\begin{align*}
q_t=&1+\frac{\gamma_0}{\lambda}\Big[\tfrac{1-e^{-\lambda t}}{2}-\tfrac{\lambda t}{2}\Big]+O(\gamma_0^2),\\
\dot q_t=&\tfrac{\gamma_0}{2}\big(e^{-\lambda t}-1\big)+O(\gamma_0^2).
\end{align*}

Then $u_\tau=(p-1)+\delta$ with $\delta=\!-\tfrac{2-p}{\lambda}(1-e^{-\lambda\tau}-\lambda\tau)\gamma_0+O(\gamma_0^2)$, and a Taylor expansion of $\text{N}_o$ around $u_\tau=p-1$ gives

\begin{equation*}
\text{N}_o=\frac{1}{2A^2}\,\delta^2+O(\gamma_0^3)=C_F(p,\lambda,\tau)\,\gamma_0^2+O(\gamma_0^3),\quad C_F>0,
\end{equation*}

\noindent while $\text{D}_o=\tfrac{(2-p)}{\sqrt{A}}\,\gamma_0\lambda\,\tau+O(\gamma_0^2)$. Hence

\begin{equation*}
\tau_{\text{\tiny QSL}}^{o}(\gamma_0)=\frac{\text{N}_o}{\text{D}_o}=C(p,\lambda,\tau)\,\gamma_0+O(\gamma_0^2),
\end{equation*}

\noindent i.e.\ it \emph{starts at zero and increases linearly} with $\gamma_0$.

\subsubsection*{Markovian regime: moderate coupling ($0<\gamma_0\le\lambda/2$)}

Here $q_t$ is real, nonnegative and decreasing; using $|q_t\dot q_t|=-\tfrac12\,\tfrac{d}{dt}(q_t^2)$ and $u=1-(2-p)q^2$ yields the \emph{exact} denominator

\begin{equation*}
\text{D}_o=\frac{2}{\tau}\Big[\operatorname{arsinh}(u_\tau)-\operatorname{arsinh}(p-1)\Big].
\end{equation*}

As $\gamma_0$ grows (for fixed $\lambda,\tau$), $q_\tau\to 0$ so $u_\tau\to 1$, and both

\begin{align*}
\text{N}_o \to &1-\frac{p}{\sqrt{2A}},\\
\text{D}_o\to &\frac{2}{\tau}\Big[\operatorname{arsinh}(1)-\operatorname{arsinh}(p-1)\Big]
\end{align*}

\emph{saturate}, producing the observed \emph{plateau} in $\tau_{\text{\tiny QSL}}^{o}$.

\subsubsection*{non-Markovian regime: strong coupling ($\gamma_0>\lambda/2$)}

Now $q_t$ oscillates with frequency $h_2$ and decays as $e^{-\lambda t/2}$. The factor $\gamma_0\lambda/h_2\sim\sqrt{\gamma_0\lambda/2}$ and the $\sin$ term boost the time–averaged $|\dot q_t|$, thus \emph{increasing $\text{D}_o$}, while oscillatory \emph{revivals} can place $\rho_\tau$ closer to $\rho_0$, so $\text{N}_o$ saturates or decreases. Consequently, $\tau_{\text{\tiny QSL}}^{o}$ decreases after the plateau (with small ripples, as shown in Figs.~\ref{fig:Foresults2Dlamb20} and~~\ref{fig:Foresults2Dlamb50}).

\subsubsection{Alternative fidelity}

From Eq. \eqref{eq:alterfidelMTQSL}, the alternative–fidelity QSL (MT-type) can be written as

\begin{align*}
\tau_{\text{\tiny QSL}}^{a}=&\frac{\text{N}_a}{\text{D}_a},
\end{align*}

where

\begin{align}
\text{N}_a &\doteq 1-\frac12\bigl(1+\tilde{g}_0 \tilde{g}_\tau\bigr)\bigl[1+\bm r\cdot\bm R_\tau\bigr]\nonumber\\[2pt]
\text{D}_a &\doteq \frac{1}{2\tau}\int_0^\tau |\dot R_z(t)|\,K(R_z(t);p)\,dt,
  \label{eq:Den_K}
\end{align}

In this case the function $\text{D}_a$ can be written as a weighted path length
along the $z$-axis of the Bloch sphere. The function $K(R_z;p)$ is nonnegative (cf. \eqref{eq:K_def}) and depends only on the Bloch
coordinate $R_z$ and the mixing parameter $p$; it is the same in all
regimes discussed below.

\subsubsection*{Markovian regime: very weak coupling
$\gamma_0\to 0^+$}

In the limit $\gamma_0\to 0^+$ (with $\lambda$ and $\tau$ fixed), the
system remains very close to its initial state for all times
$t\le\tau$. In this regime $\mathcal{F}_a(\rho_0,\rho_\tau)$ admits the expansion
$\mathcal{F}_a(\rho_0,\rho_\tau)
= 1 - C(p,\tau)\,\gamma_0^2 + \mathcal{O}(\gamma_0^3)$, so that
$\text{N}_a \sim \gamma_0^2$ as $\gamma_0\to 0^+$. At the same
time, the evolution of the Bloch coordinate $R_z(t)$ is very slow and
$R_z(t)\simeq r_z$ for all $t\in[0,\tau]$. As a consequence, one can
approximate the weight in Eq.~\eqref{eq:Den_K} by
$K(R_z(t);p)\simeq K(r_z;p)$, with

\begin{align*}
  K(r_z;p)
  &= g_0\,g(r_z)^{-1}\,
     \frac{2\,\bigl|r_z\,[1+r_z^2]\bigr|}
          {\bigl[1+r_z^2\bigr]^2}\nonumber\\[2pt]
   &+ \sqrt{1+r_z^2}
   + g(r_z)\,\sqrt{1-r_z^2},
\end{align*}

\noindent and $|\dot R_z(t)|\propto\gamma_0$, implying $\text{D}_a\sim
\gamma_0$. Hence,

\begin{equation*}
  \tau_{\mathrm{QSL}}^a(\gamma_0)
  = \frac{\text{N}_a}{\text{D}_a}
  \sim \gamma_0,\qquad \gamma_0\to 0^+.
\end{equation*}

Thus, the quantum speed limit starts from zero and increases
approximately linearly with the coupling strength in the very
weak-coupling Markovian regime. In this limit the function $K(R_z;p)$
is effectively evaluated at the initial Bloch point $R_z=r_z$.

\subsubsection*{Markovian regime: moderate coupling
$0<\gamma_0\lesssim\lambda/2$}

For moderate Markovian couplings ($\gamma_0\le\lambda/2$ and
$\gamma_0\tau\gtrsim 1$), the excited-state population decays
significantly within the interval $[0,\tau]$, and the evolution of
the Bloch coordinate is well approximated by
$R_z(t)=1-(2-p)e^{-\Gamma t}$, with an effective decay rate
$\Gamma\propto\gamma_0$. As $\gamma_0$ increases in this window, the
final value $R_z(\tau)$ rapidly approaches $1$, so that
$\rho_\tau\simeq\ket{0}\!\bra{0}$ and $\mathcal{F}_a(\rho_0,\rho_\tau)$
saturates to
$\mathcal{F}_a(\rho_0,\rho_\tau)\to\Tr(\rho_0\ket{0}\!\bra{0})
= p/2$. Hence, the numerator approaches the constant

\begin{equation}
  \text{N}_a(p)
  \simeq 1 - \frac{p}{2}.
\label{eq:Na(p)plateau}
\end{equation}

In the same Markovian approximation, the monotonicity of $R_z(t)$
allows us to change variables in Eq.~\eqref{eq:Den_K} from $t$ to
$R_z=R_z(t)$, which yields

\begin{equation}
  \text{D}_a(\gamma_0)
  = \frac{1}{2\tau}\int_{R_z=p-1}^{R_z=R_z(\tau)}K(R_z;p)\,dR_z,
  \label{eq:Den_Rz_K}
\end{equation}

\noindent with $K(R_z;p)$ defined in Eq.~\eqref{eq:K_def}. As
$R_z(\tau)\to 1$ for increasing $\gamma_0$, the integral in
Eq.~\eqref{eq:Den_Rz_K} saturates to the purely geometric quantity

\begin{equation*}
  \text{D}_a(p,\tau)
  \simeq \frac{1}{2\tau}\int_{p-1}^{1}K(R_z;p)\,dR_z,
\end{equation*}

\noindent which is independent of $\gamma_0$ (within the Markovian
approximation). Consequently, in this intermediate regime both the
distance encoded in $\text{N}_a$ and the path length encoded in
$\text{D}_a$ become nearly coupling-independent, and

\begin{equation*}
  \tau_{\mathrm{QSL}}^a(\gamma_0)
  \simeq \tau_{\mathrm{QSL}}^a(p,\tau)
  = \frac{2\tau\left(1-\tfrac{p}{2}\right)}
         {\displaystyle\int_{p-1}^{1}K(R_z;p)\,dR_z},
\end{equation*}

\noindent which appears as the plateau region in the numerical plots. Here the
function $K(R_z;p)$ weights the different points along the straight
segment in Bloch space connecting the initial Bloch coordinate
$R_z=p-1$ with the ground state $R_z=1$.

\subsubsection*{Non-Markovian regime: strong coupling
$\gamma_0 > \lambda/2$}

When $\gamma_0>\lambda/2$, the dynamics enter the non-Markovian
regime and the amplitude $q_t$ acquires oscillatory contributions.
Accordingly, $R_z(t)=1-(2-p)q_t^2$ becomes oscillatory as a function
of time, with a number of oscillations within $[0,\tau]$ that
increases with $\gamma_0$. In the parameter range considered here,
the endpoint $R_z(\tau)$ remains close to $1$ even for strong
couplings, so that the numerator $\text{N}_a$ stays approximately
equal to its plateau value $\text{N}_a(p)$ (cf.~\eqref{eq:Na(p)plateau}).

In contrast, the denominator retains the path-length structure
\begin{equation*}
  \text{D}_a(\gamma_0)
  = \frac{1}{2\tau}\int_0^\tau |\dot R_z(t)|
    K\bigl(R_z(t);p\bigr)\,dt,
\end{equation*}

\noindent with the same weight $K(R_z;p)$ as in the Markovian analysis. In the
non-Markovian regime, every oscillation of $R_z(t)$ adds an extra
positive contribution to the total variation
$\int_0^\tau|\dot R_z(t)|\,dt$, and therefore to
$\text{D}_a(\gamma_0)$. As $\gamma_0$ increases, the total path
length grows, while the endpoint-based distance $\text{N}_a$
remains essentially constant. Consequently,

\begin{equation*}
  \tau_{\mathrm{QSL}}^a(\gamma_0)
  = \frac{\text{N}_a(\gamma_0)}{\text{D}_a(\gamma_0)}
  \quad\text{decreases for }\gamma_0>\lambda/2,
\end{equation*}

\noindent in agreement with the numerical observation that the plateau region in
$\tau_{\mathrm{QSL}}^a$ is followed by a decay in the strongly
non-Markovian regime. The same local weight $K(R_z;p)$ governs all
three regimes; what changes with $\gamma_0$ is the geometry of the
trajectory $R_z(t)$ through the Bloch space.

\subsubsection{Comparative QSLs analysis}

For a pure initial state of the form $\rho_0=\ket{1}\!\bra{1}$, the ML-type QSL $\tau_{\text{\tiny QSL}}^B$ built with the operator norm is \emph{tight} in the Markovian regime, since on that interval it coincides with the driving time $\tau$ (here, $\tau=1$ in the referenced figure). In the non-Markovian regime, i.e., in the strong-coupling domain $\gamma_0>\lambda/2$, non-Markovian effects can increase the time–averaged generator norm (information backflow), and consequently the value of $\tau_{\text{\tiny QSL}}^B$ drops below its Markovian value. These observations are in complete agreement with the results reported in~\cite{Deffner2013}.\par
The  \emph{qualitative} behavior of $\tau_{\text{\tiny QSL}}^s$, $\tau_{\text{\tiny QSL}}^o$, and $\tau_{\text{\tiny QSL}}^a$ curves across the different coupling regimes determined by $\gamma_0$ is similar. For $0<p\le 1$, the QSL curves exhibit an approximately linear increase for $0<\gamma_0\ll\lambda$, followed by a plateau for moderate couplings $\gamma_0<\lambda/2$, and subsequently a decrease for $\gamma_0>\lambda/2$, over the ranges displayed in the figures. These trends can be explained from the closed forms derived for the various QSLs. As expected, for $\tau_{\text{\tiny QSL}}^s$ with $0<p\le 1$, the tightest QSL is obtained when the operator norm is used. Moreover, since super fidelity $\mathcal{F}_s(\rho_0,\rho_t)$ reduces to the Bures fidelity $F_B(\rho_0,\rho_t)$ when the initial state is pure, i.e., $\rho_0=\ket{1}\!\bra{1}$ in the scenario analyzed, the $\tau_{\text{\tiny QSL}}^s$ curve computed with $\mathcal{F}_s(\rho_0,\rho_t)$ and the operator norm \emph{coincides} with the curve for $\tau_{\text{\tiny QSL}}^{B}$ (with Bures fidelity) with the operator norm; see Figs.~\ref{fig:FBresults2Dlamb20} and \ref{fig:FBresults2Dlamb50}, as anticipated.\par
For mixed initial states of the form $\rho_0=\tfrac{p}{2}\mathbb I+(1-p)\ket{1}\!\bra{1}$, the QSLs $\tau_{\text{\tiny QSL}}^o$, and $\tau_{\text{\tiny QSL}}^a$ are looser than $\tau_{\text{\tiny QSL}}^s$ when the operator norm is used. This behavior is expected, because $\tau_{\text{\tiny QSL}}^{o}$ and $\tau_{\text{\tiny QSL}}^{a}$ are MT–type QSLs and, in general, are not tight. Since $\mathcal{F}_a(\rho_0,\rho_t)$ coincides with the Bures fidelity $F_B(\rho_0,\rho_t)$ when the initial state is pure (that is, when $p=0$ and $\rho_0=\ket{1}\!\bra{1}$ in this scenario), the QSL $\tau_{\text{\tiny QSL}}^{a}$ reduces to the MT–type Bures QSL $\tau_{\text{\tiny QSL}}^{B}$ when the Hilbert--Schmidt norm is employed, as expected. By contrast, $\mathcal{F}_o(\rho_0,\rho_t)$ does \emph{not} reduce to $F_B(\rho_0,\rho_t)$ for pure $\rho_0$ due to the additional factor $1/\sqrt{\Tr(\rho_t^2)}$ in $F_o$ (cf. Eq. \eqref{eq:operfidel}); therefore, for $p=0$ one has that $\tau_{\text{\tiny QSL}}^{o}$ does \emph{not} reduce to $\tau_{\text{\tiny QSL}}^{B}$ when the Hilbert--Schmidt norm is used, again as expected.\par
Unlike the results presented in Ref.~\cite{Deffner2013}---which strictly applies when the initial state $\rho_0$ is pure---the use of generalized fidelities such as $\mathcal{F}_s(\rho_0,\rho_t)$, $\mathcal{F}_o(\rho_0,\rho_t)$, and $\mathcal{F}_a(\rho_0,\rho_t)$ allows us to propose QSLs that are applicable when the initial state $\rho_0$ is \emph{mixed}. Although the QSLs derived from $\mathcal{F}_s(\rho_0,\rho_t)$, $\mathcal{F}_o(\rho_0,\rho_t)$, and $\mathcal{F}_a(\rho_0,\rho_t)$ are in general not tight, they provide nontrivial constraints on the evolution time in the mixed-state setting, which constitutes a meaningful extension of the geometric approach explored here. All of our qualitative conclusions are fully consistent with Ref.~\cite{Deffner2013}, and with subsequent analyses of QSLs based on $\mathcal{F}_s(\rho_0,\rho_t)$, $\mathcal{F}_o(\rho_0,\rho_t)$, and $\mathcal{F}_a(\rho_0,\rho_t)$ for mixed initial states of the form $\rho_0=\tfrac{p}{2}\mathbb I+(1-p)\ket{+}\!\bra{+}$ with $\ket{+}=(\ket{0}+\ket{1})/\sqrt{2}$ (see, e.g., Refs.~\cite{Wu2020,Sun2015,Ektesabi2017}).\par

\section{Concluding remarks}
\label{sec:concludingremarks}

We have analyzed the role of functionals of generalized fidelity measures in the derivation of quantum speed limits (QSLs) within a geometric framework. We have demonstrated that any monotone and differentiable reparametrization of the selected fidelity produces identical ML/MT-type bounds, thereby rendering the QSL invariant under such transformations and providing a unified perspective connecting fidelity- and metric-induced geometric formulations. This invariance clarifies the inherent limitations of attempts to improve QSLs through functional transformations of fidelity and indicates that meaningful advances must instead arise from alternative fidelity definitions. The general structure developed in this work reveals that, once a particular generalized fidelity is specified, the corresponding ML- and MT-type bounds for both unitary and nonunitary (Lindblad-type) dynamics depend exclusively on that choice of fidelity measure. Furthermore, we have shown that various QSLs previously reported in the literature naturally emerge as particular instances of our framework (e.g.,~\cite{Deffner2013,Wu2020,Sun2015,Ektesabi2017,Wu2018}), suggesting a coherent and general foundation upon which further developments—based on other generalized fidelity measures—may be constructed. Moreover, the results presented here can be straightforwardly extended by considering generalized fidelity measures beyond those employed as illustrative examples in the present study.\par
Using the damped Jaynes--Cummings (DJC) model as a concrete physical setting, we verified that, for a \emph{pure} initial state $\rho_0=\ket{1}\!\bra{1}$ (as in Ref.~\cite{Deffner2013}), the Bures fidelity-based ML-type QSL with the operator norm is \emph{tight} throughout the Markovian regime ($\gamma_0\le \lambda/2$), yielding a plateau at the actual driving time $\tau$. Upon entering the non-Markovian regime ($\gamma_0>\lambda/2$), the time dependence of the decay rate---interpretable as information backflow---increases the time-averaged generator norm, so the Bures fidelity-based ML-type QSL decreases below its Markovian value; this plateau-and-decrease trend is fully consistent with the results presented in Ref.~\cite{Deffner2013} (cf. Sec.~\ref{ssec:buresfidelity}). For the generalized fidelities $\mathcal{F}_s(\rho_0,\rho_t)$ (super-fidelity, cf. Eq.~\eqref{eq:superfidelity}), $\mathcal{F}_o(\rho_0,\rho_t)$ (operator fidelity as defined here, cf. Eq.~\eqref{eq:operfidel}), and $\mathcal{F}_a(\rho_0,\rho_t)$ (alternative fidelity, cf. Eq.~\eqref{eq:ektesabifidelity}), the qualitative dependence of the QSL on the coupling strength is similar across regimes: an approximately linear rise for $0<\gamma_0\ll\lambda$, a plateau for moderate coupling $\gamma_0<\lambda/2$, and a decrease for $\gamma_0>\lambda/2$ (cf. Secs.~\ref{ssec:Super-fidelity}, \ref{ssec:operatorfidelity}, and \ref{ssec:alterfidelity}). In the case of $\mathcal{F}_s(\rho_0,\rho_t)$ and for $0<p\le 1$, the operator norm yields the tightest QSLs; moreover, since $\mathcal{F}_s(\rho_0,\rho_t)$ coincides with the Bures fidelity for pure $\rho_0$, the QSL obtained from $\mathcal{F}_s(\rho_0,\rho_t)$ with the operator norm reproduces the Bures fidelity-based ML-type QSL with the operator norm when $\rho_0=\ket{1}\!\bra{1}$, in complete agreement with~\cite{Deffner2013}. For mixed initial states of the form $\rho_0=\tfrac{p}{2}\mathbb I+(1-p)\ket{1}\!\!\bra{1}$, the QSLs derived from $\mathcal{F}_o(\rho_0,\rho_t)$ and $\mathcal{F}_a(\rho_0,\rho_t)$ are looser than those from $\mathcal{F}_s(\rho_0,\rho_t)$ under the operator norm, in line with their MT-type character and the general non-tightness of such QSLs. Moreover, our results indicate that the MT-type QSLs obtained by means of $\mathcal{F}_o(\rho_0,\rho_t)$ are looser than the corresponding MT-type QSLs derived from $\mathcal{F}_a(\rho_0,\rho_t)$, in complete agreement with Ref.~\cite{Ektesabi2017}.
Crucially, while the Bures fidelity-based QSLs are directly applicable to pure initial states, the use of $\mathcal{F}_s(\rho_0,\rho_t)$, $\mathcal{F}_o(\rho_0,\rho_t)$, and $\mathcal{F}_a(\rho_0,\rho_t)$ extends the geometric QSL approach analyzed in this work to mixed initial states, providing nontrivial---albeit generally non-tight---QSLs.\par

\bigskip
\section*{Acknowledgments}
T.M.O, M. P. and P.W.L are fellows of the National Research Council of Argentina (CONICET). Y. A. has a fellowship from CONICET.
The authors are grateful to CONICET (PIP N$^\circ$: 135/23), Secretaria CyT-UNLP (Proyecto 11/X959), and Secretaria de Ciencia y T\'ecnica de la Universidad Nacional de C\'ordoba (SeCyT-UNC, Argentina, Proyecto Consolidar 2023-2025) for financial support.

\end{document}